\documentclass[nohyper]{JHEP3}
\usepackage{amssymb}

\newcommand{\bmat}{\left(\begin{array}}
\newcommand{\emat}{\end{array}\right)}

\def\yzero{\smash{\hbox{$y\kern-4pt\raise1pt\hbox{${}^\circ$}$}}}

\def\-{\hphantom{-}}

\def\beq{\begin{equation}}
\def\eeq{\end{equation}}
\def\ber{\begin{eqnarray}}
\def\eer{\end{eqnarray}}
\def\s2{\frac{1}{2}}

\def\beq{\begin{equation}}
\def\eeq{\end{equation}}
\def\beqa{\begin{eqnarray}}
\def\eeqa{\end{eqnarray}}

\def\IF{\relax{\rm I\kern-.18em F}}
\def\II{\relax{\rm I\kern-.18em I}}
\def\IP{\relax{\rm I\kern-.18em P}}
\def\IC{\relax\hbox{\kern.25em$\inbar\kern-.3em{\rm C}$}}
\def\IR{\relax{\rm I\kern-.18em R}}

\def\Dsl{\,\raise.15ex\hbox{/}\mkern-13.5mu D} 

\def\IC{\bf C}

\title{Free field realization of superstring theory on $AdS_3$}
\author{Diego M. Hofman and Carmen A. N\'u\~nez\\ Instituto de Astronom\'{\i}a y
F\'{\i}sica del Espacio (IAFE)\\
C.C. 67 - Suc. 28, 1428 Buenos Aires, Argentina\\
and\\
Physics Department, University of Buenos Aires\\ E-mail:
\email{dhofman@iafe.uba.ar}, \email{carmen@iafe.uba.ar}}

\abstract{The Coulomb gas representation of expectation values in
$SU(2)$ conformal field theory developed by Dotsenko is extended
to the $SL(2,\mathbb R)$ WZW model and applied to bosonic string
theory on $AdS_3$ and to Type II superstrings on $AdS_3\times
{\cal {N}}$. The spectral flow symmetry is included in the free
field realization of vertex operators creating superstring states
of both Ramond and Neveu-Schwarz sectors. Conjugate
representations for these operators are constructed and a
background charge prescription is designed to compute correlation
functions. Two and three point functions of bosonic and fermionic
string states in arbitrary winding sectors are calculated.
Scattering amplitudes that violate winding number conservation are
also discussed.}


\begin{document}

\section{Introduction}

Two and three dimensional toy models of string theory have been useful to
explore some essential features of theoretical physics in a setting with a
vastly reduced number of dynamical degrees of freedom. Particularly
interesting examples can be found in nonperturbative physics, the
continuation to Lorentzian
signature, the notion of time in curved backgrounds,
 singularities
 and conceptual problems of black
hole physics.

One of the simplest frames where these
questions can be addressed is the $SL(2, \mathbb R)$ group manifold.
The WZW model on $SL(2, \mathbb R)$ is an exact conformal field theory
describing string propagation in three
dimensional Anti de Sitter spacetime ($AdS_3$) \cite{balog}. This model is
closely related to
two and
three
dimensional black holes in string theory through the $SL(2,
\mathbb R)/U(1)$ coset
\cite{witten} and orbifolding
\cite{btz}  respectively.
$SL(2, \mathbb R)$ cosets are also linked to Liouville theory
and some of its generalizations
\cite{polyakov, seiberg} and to the physics of defects or singularities
\cite{ov, litstring}.

Despite its simplicity, the efforts to develop a consistent string theory
on $AdS_3$ turned out to be highly non trivial.
The origin of the difficulties can be traced to the
non-compact nature of $SL(2,\mathbb R)$
and the non-rational structure of the worldsheet CFT
\cite{todos}. The resolution of the main problem, the apparent lack of
unitarity of the theory, was possible with the help of the
$AdS$/CFT duality conjecture \cite{malda}.
The impact of the conjecture was twofold:
on the one hand this is an example where the $AdS$/CFT duality
has been explored beyond the
supergravity
approximation, with complete control over the theory in the bulk
\cite{gks, ks}; on the other hand, the conjecture provided
a productive feedback on the interpretation of the puzzles raised by the
worldsheet theory \cite{mo3}.

The structure of the Hilbert space of the $SL(2, \mathbb R)$ WZW model was
 determined in \cite{mo1} where the spectrum of physical states of
string theory on $AdS_3$ and a proof of the no-ghost theorem were given
as well.
It was realized that the model has a spectral flow symmetry which
gives rise to new representations for the string spectrum besides
the standard discrete and continuous unitary series which had been
considered previously \cite{todos}. The computation of the one
loop partition function
performed in  \cite{mo2, ikp}, provided further evidence for the
spectrum of the free theory.

To establish the consistency of the full theory one has to consider
interactions and verify the closure of the operator product
expansion. But
the fusion rules are difficult
to find  in the non-compact worldsheet CFT that defines string theory on
$AdS_3$ because there are generically no null
vectors in the relevant current
algebra representations, so that most of the techniques from rational
conformal field theories are not available and consequently the
factorization
properties of the model have not been completely determined yet.
Nevertheless,
important progress has been achieved recently in the resolution of this
problem. Correlation functions of primary fields have been calculated
using different procedures in the Euclidean version of the theory, the
$\frac {SL(2,
\mathbb C)}{SU(2)}\equiv H_3^+$ model. The path integral method to obtain
expectation values was
started in \cite{gawe} and applied to the computation of two and three
point amplitudes of bosonic string states in \cite{ishi}.
A generalization of the bootstrap
approach was designed by Teschner and some four
point functions were given in \cite{teschner}.
The
 physical interpretation of these exact results was performed by
Maldacena and Ooguri in \cite{mo3} where correlation functions involving
spectral flowed operators were also presented.

Scattering amplitudes of $n-$ states in string theory on $AdS_3$
exhibit several subtleties for $n\ge 3$. On the one hand, correlation
functions of discrete states are only well defined if
the sum of the isospins $j$ of the external operators satisfies
$\sum_i j_i < k -3$ (where $k$ is the level of the Kac Moody algebra).
Moreover
the four point functions do not factorize as expected into a sum of
products of three point functions with physical intermediate states unless
the quantum numbers of the external states verify
$j_1+j_2<\frac {k-3}2$ 
and $j_3+j_4<\frac {k-3}2$.
The meaning of
these constraints was proposed in  \cite{mo3}: correlation functions
violating these bounds do not represent a well-defined computation in the
dual CFT$_2$. However
 one would like to better understand this unusual feature from the
worldsheet
viewpoint.

On the other hand, a curious aspect of this model is that physical
amplitudes of $n$ string states
 may violate winding number conservation up to $n-2$ units. This fact
is well
understood
from the representation theory of $SL(2, \mathbb R)$ \cite{mo3}.
Nevertheless the computation of winding non-conserving
scattering amplitudes
proposed in \cite{mo3}
involves the insertion of
spectral flow operators in the correlation functions. This implies the
computation of expectation values
of more vertex operators than the $n$ original ones.
This procedure has been applied to
three point functions, but four point functions
violating winding number conservation by one or two units require the
calculation of correlators
with
five or six operator insertions, with the consequent complications.
These amplitudes
 are needed
to definitely establish the unitarity of the theory through the analysis
of their
factorization properties.

Therefore it seems  necessary to develop techniques that simplify
these computations and allow to perform others that would clarify the full
structure of the model.
The free field description of the theory appears as a powerful tool in
this direction.
This approximation was initially applied in \cite{gks, ks} to derive the
spacetime CFT and establish the conjectured $AdS$/CFT correspondence in
the three dimensional case (for related work see \cite{gknotes}).
Even though this approach is expected to give a good picture of the theory
only near the boundary of $AdS_3$,
the computation of two and three point amplitudes of string states
using the Coulomb gas formalism in \cite{gn2, gn3} has produced results in
complete agreement with the exact ones. Moreover, the analysis of
unitarity
in this approximation might give important information on the
consistency of the complete
theory. For that reason, the aim of this article
is to further develop this approach and extend the formulation to the
supersymmetric case.

The Coulomb gas formalism to compute  conformal blocks in the
compact $SU(2)$ CFT
was
introduced by Dotsenko and applied to the computation of conformal blocks
 for integer $2j$ and admissible representations
in \cite{dotse}.
The non-compact
$SL(2, \mathbb R)$ case was considered in \cite{andreev, fgp} where
the degenerate case was resolved.
Two and
three point amplitudes of physical string states in the coset
theory $SL(2, \mathbb R)/U(1)$ were computed in \cite{becker}.
More recently it was extended to take into account the
spectral flow symmetry of $SL(2, \mathbb R)$ in \cite{gn2} and applied  to
the computation of
two and three point
functions of $AdS_3$ string states  for
arbitrary winding
sectors, both preserving and violating winding number conservation, in
\cite{gn3}.

Here we study
superstring theory on $AdS_3 \times {\cal N}$. We consider vertex
operators
 both
in the Ramond and Neveu-Schwarz sectors in the free field approximation
and construct their conjugate
representations. We develop Dotsenko's background charge prescription
 to compute  expectation values in the supersymmetric theory
and employ it to calculate
scattering amplitudes of two and three
superstring
states. We analyze the structure of winding (non)conservation pattern and
discuss the relevance of the internal theory concerning this question.

The organization of the paper is as follows.
In Section 2 the Coulomb gas representation of
correlators in bosonic string theory on
$AdS_3$
is reviewed and completed. In Section 3 the realization of the
supersymmetric theory is presented and correlation functions of two and
three superstring states are computed, both in the Ramond and Neveu
Schwarz sectors of the theory. Conclusions are presented in Section 4.

\section{Bosonic string theory on $AdS_3$}

In this section we briefly review string theory on $AdS_3$ mainly to
introduce
our notation and conventions. We complete the Coulomb gas formulation
of expectation values of string states
started in \cite{gn2, gn3}.

The  metric of Euclidean $AdS_3$ (the hyperbolic space $H_3^+$) 
can be written in {\it Poincar\'e}
coordinates as
\beq
ds^2 = l^2(d\phi^2 + e^{2\phi}d\gamma d\bar\gamma) \quad .
\label{h3}
\eeq

\noindent
where $\phi\in\mathbb R$, $\{\gamma , \bar\gamma \}$ are complex
coordinates parametrizing
the boundary of $H^+_3$ which is located at $\phi\rightarrow\infty$ and
the parameter $l$ is related to the scalar curvature as ${\cal R}=-2/l^2$.

Consistent string propagation in this background metric requires
in addition an antisymmetric rank two tensor background field
$B = l^2 e^{2\phi}d\gamma \wedge d\bar
\gamma$. The theory
 is described by a non linear sigma model with action

\beq
S = \frac k{8\pi} \int d^2 z (\partial\phi \bar\partial\phi + e^{2\phi}
\bar\partial\gamma  \partial\bar\gamma)
\label{sigmod}
\eeq

\noindent where $k=l^2/l^2_s$ and $l^2_s$ is the fundamental string
length, which is equivalent to a WZW model on $SL(2,\mathbb R)$ (or
actually its Euclidean version $SL(2,\mathbb C)/SU(2)$). 
This action has a larger
symmetry than the isometries of the group, namely
$g(z,\bar z) \rightarrow \Omega (z) g(z, \bar z) \bar \Omega ^{-1}(\bar
z)$,
with $g, \Omega \in SL(2, \mathbb R)$. The corresponding currents
$J(z) = - \frac k2 (\partial g) g^{-1},
\bar J( \bar z) = - \frac k2 (\bar\partial g ^{-1}) g$
can be expanded in Laurent series 
\beq
J^a(z) = \sum_{n=-\infty}^\infty J_n^a ~z^{-n-1}
\eeq

\noindent
and the coefficients $J^a_n$ satisfy a Kac-Moody algebra given by

\beq
[J^a_n, J^b_m] = i\epsilon^{ab}_c J^c_{n+m} - {k\over 2}\eta^{ab} n
\delta_{n+m, 0} \label{algebra} \quad ,
\eeq

\noindent
where the Cartan Killing metric is $\eta^{+-} = \eta^{-+} = 2$,
$\eta^{33}=-1$
and $\epsilon^{ab}_{c}$ is the Levi Civita antisymmetric tensor.
And similarly for the antiholomorphic currents.

The Sugawara stress-energy tensor  is given by

\beq
T = \frac {\eta_{ab}}{k-2} :J^a(z)J^b(z):
\label{suga}
\eeq

\noindent
It is related to the Casimir of the group as $C = (k-2) T$
and it leads to the following central charge of the Virasoro algebra

\beq
c= {3k\over k-2} =
3 + {12\over \alpha_+^2} \quad ,
\eeq

\noindent
($\alpha_+ = \sqrt{2(k-2)}$).

The classical solutions of this theory were presented in
\cite{mo1}. Timelike geodesics oscilate around the center of
$AdS_3$ whereas spacelike geodesics representing tachyons travel
from one side of the boundary to the opposite. Solutions
describing string propagation are obtained from the dynamics of
pointlike particles through the spectral flow operation. Timelike
geodesics give rise to {\it short strings}, bound states trapped
in the gravitational potential of $AdS_3$. Conversely, {\it long
strings} arising from spacelike geodesics can reach the boundary
of $AdS_3$. The spectral flow parameter $w$ is an integer named {\it
winding number}.
Different values of $w$ correspond to distinct solutions, even at
the classical level (as exhibited, for instance, by the energy
spectrum).

At the quantum level, the building blocks of the Hilbert space ${\cal H}$
are unitary hermitic representations of $SL(2, \mathbb R)$.
The states $|j, m>$ satisfy
\beqa
C_0 |j, m> = j(j+1) |j,m> \quad , && \quad J_0^3 |j,m > = m |j, m > ,
\nonumber \\
J_0^\pm |j, m \rangle & = & (m \mp j)|j, m\pm 1 \rangle \quad , \eeqa
\noindent with \beq \{ m \in \mathbb{R} , j \in \mathbb{R} \}
\quad \vee \quad \{ m \in \mathbb{R} , j \in - \frac{1}{2} + i
\mathbb{R} \} \eeq

\noindent
as required by hermiticity and in addition they must be Kac Moody
primaries, namely
\beq
J^a_n |j,m \rangle = 0
\quad \forall ~ n > 0 \quad .
\eeq

The allowed representations are:

$\bullet$
 Discrete lowest weight representation
\beq
{\cal D}^+_j=\{ |j,m \rangle ; ~~j \in  \mathbb R ; ~~ m = j+1, j+2,
j+3,...\}
\eeq

$\bullet$  Discrete highest weight representation
\beq
{\cal D}^-_j=\{ |j,m \rangle ; ~~ j \in \mathbb R ; ~~ m = -j-1, -j-2,
-j-3,...\} \label{dj+}
\eeq

$\bullet$ Principal continuous representation
\beq
{\cal C}_j^\alpha =
\{|j, m \rangle ; ~~j = -\frac 12 + i \lambda ; ~\lambda \in \mathbb R ;~~
m = \alpha , \alpha \pm 1, \alpha \pm 2, ... ; ~~\alpha \in \mathbb R\}
\label{dj-}
\eeq 

For applications to string theory
one  considers the universal
cover of $SL(2,\mathbb R)$, where $j$ is not quantized. Notice that the
vectors in
${\cal H}$ related by $j \leftrightarrow -1-j$ represent the same physical
state and therefore $j$ can be restricted to $j \ge - \frac 12$.
 The complete basis of
${\cal L}^2(AdS_3)$ is given by
 ${\cal
C}^\alpha_{j=-1/2+i\lambda}\times{\cal C}^\alpha_{j=-1/2+i\lambda}$ and
${\cal
D}_j^\pm \times{\cal D}_j^\pm$ with $j > - 1/2$.

The representation space can be enlarged by acting on
the primary states in
these series with $J_{n}^{a}$, $n<0$.
The corresponding
representations are denoted by ${\widehat{\cal D}_j^{\pm }},
\widehat{\cal C}_{j}^\alpha$.
Furthermore the full representation space contains the spectral flow
images of these series which correspond to winding classical strings.
Actually the spectral flow operation leads to the following automorphism
of the $SL(2, \mathbb R)$ currents
\beqa
J_n^3 & \rightarrow & \widetilde J_n^3 = J_n^3 - \frac k2 w \delta_{n,0}
\label{j3tilde}\\
J_n^\pm & \rightarrow & \widetilde J_n^\pm = J_{n\pm w}^\pm
\eeqa

\noindent
with $w \in \mathbb Z$ and consequently the modes of the Virasoro
generators transform as
\beq
L_n \rightarrow \widetilde L_n = L_n + w J_n^3 - \frac{k}{4} {w}^2
\delta_{n,0} \quad .
\eeq

Unlike the compact $SU(2)$ case, the new operators generate inequivalent
representations of $SL(2, \mathbb R)$ with states $|\tilde j , \tilde m ,
\omega \rangle$ satisfying
\beq
\widetilde{L}_0 | \widetilde{j} , \widetilde{m} , w \rangle =
-\frac{\widetilde{j}(\widetilde{j}+1)}{k-2} | \widetilde{j}
,\widetilde{m} , w \rangle \quad , \quad
\widetilde{J}_0^3 | \widetilde{j} , \widetilde{m} , w \rangle =
\widetilde{m} | \widetilde{j} , \widetilde{m} , w \rangle \label{jmw}
\eeq

Finally, the complete Hilbert space of string theory on $AdS_3$ is
obtained by
applying creation operators $ \widetilde J^a_n, n<0$ on the primary states
defined by
(\ref{jmw}) and verifying the physical state conditions
\begin{eqnarray}
(L_0 - 1) | \widetilde{j} , \widetilde{m} , w , \widetilde{N} , h
\rangle &=& \left(- \frac{\widetilde{j}(\widetilde{j}+1)}{k-2} - w
\widetilde{m} - \frac{k}{4} w^2
 + \widetilde{N} + h -1 \right) | \widetilde{j}
,\widetilde{m} , w , \widetilde{N} , h \rangle = 0 \nonumber \\
\label{vjmw} \\
L_n | \widetilde{j} , \widetilde{m} , w , \widetilde{N} , h
\rangle &=& \left( \widetilde{L}_n - w \widetilde{J}^3_n \right) |
\widetilde{j} , \widetilde{m} , w , \widetilde{N} , h \rangle = 0
\quad \textrm{for} ~ \, n > 0 \quad \,\, \label{vnjmw}
\end{eqnarray}

\noindent
where $\widetilde N$ is the excitation level of $\widetilde J_n$ and $h$
is
the
conformal weight of the state in the internal theory \footnote{We have
been considering string theory on $AdS_3$, but more generally we could
take a background $AdS_3 \times {\cal N}$, with ${\cal N}$ a compact
internal manifold. An interesting example has been considered
recently in relation to string amplitudes in the plane wave limit
of $AdS_3\times S^3$ in \cite{kiri}.}.

Notice that the representations
$\widehat {\cal D}_{\tilde j}^{\pm ,
w = \mp 1}$ and
$\widehat {\cal D}_{\frac k2 - 2 - \tilde j}^{\mp, w = 0}$ are
equivalent. This has an important consequence on the values allowed
for $j$. Indeed, recalling the symmetry $j \leftrightarrow -1-j$ which
implies $j \ge -\frac 12$, $j$ is restricted as required by
the no-ghost theorem \cite{mo1} to
\beq
-\frac 12 < j < \frac {k-3}2
\eeq

 \subsection{Free field representation of string theory on $AdS_3$}

The free field formulation of this theory follows from the action
(\ref{sigmod}) which can be rewritten as a free field model by
introducing auxiliary fields $\beta ,
\bar\beta$ as

\begin{equation} \label{slibre2}
S = \frac{k}{8\pi} \int \textrm{d}^2 z (\partial \phi
\overline{\partial} \phi + \beta \overline{\partial} \gamma +
\overline{\beta}
\partial \overline{\gamma} - \beta \overline{\beta} e^{-2 \phi} ) \quad .
\end{equation}

Quantization leads to include some renormalization factors \cite{david} as

\begin{equation} \label{slibre3}
S =  \frac{1}{4\pi} \int \textrm{d}^2 z (\partial \phi
\overline{\partial} \phi - \frac{2}{\alpha_+} R \phi + \beta
\overline{\partial} \gamma + \overline{\beta}
\partial \overline{\gamma} - \beta \overline{\beta}
e^{-\frac{2}{\alpha_+} \phi} )
\end{equation}

\noindent
where $R$ is the scalar curvature of the worldsheet. The interaction term
$\beta\bar\beta e^{-\frac 2{\alpha_+}\phi}$ becomes negligible near the
boundary ($\phi\rightarrow \infty$) and the theory can thus can be treated
perturbatively in this region. It can be fully described in terms of OPEs
of free
fields, namely
\beq
\phi(z) \phi(z^\prime) \sim -\ln{(z-z^\prime)} \qquad , \qquad
\gamma(z) \beta(z^\prime) \sim - \frac{1}{z-z^\prime} \quad .
\label{OPE}
\eeq

The currents are defined in the
\textit{Wakimoto} representation \cite{waki} as
\begin{eqnarray}
J^+(z) &\equiv& -\beta(z) \label{j+bg}\\
J^3(z) &\equiv& - \beta(z) \gamma(z) - \frac{\alpha_+}{2} \partial
\phi(z) \label{j3bg}\\
J^-(z) &\equiv& -\beta(z) \gamma^2(z) - \alpha_+ \gamma(z)
\partial \phi(z) - k \partial \gamma(z) \label{j-bg}
\end{eqnarray}

\noindent
and the energy momentum tensor is

\begin{equation} \label{Tlibre}
T(z) = \beta(z) \partial \gamma(z) - \frac{1}{2} \partial
\phi(z)
\partial \phi(z) - \frac{1}{\alpha_+} \partial^2 \phi(z) \quad .
\end{equation}

It is easy to see that the $\beta\gamma$ fields form a commuting
$bc-$system with conformal
weight $1$ ($\beta$) and $0$ ($\gamma$) and ghost charge $1$.

The currents satisfy the OPEs
\begin{eqnarray}
J^+(z) J^-(z^\prime) &\sim& \frac{k}{(z-z^\prime)^2} - \frac{2 J^3
(z^\prime)}{z-z^\prime}
\label{OPE+-}\\
J^3(z) J^\pm(z^\prime) &\sim& \pm \frac{J^\pm(z^\prime)}{z-z^\prime}
\label{OPE3+-}\\
J^3(z) J^3(z^\prime) &\sim& - \frac{\frac{k}{2}}{(z-z^\prime)^2} \label{OPE33}
\end{eqnarray}

\noindent
in full agreement with the commutation relations (\ref{algebra}).

\subsection{Vertex operators}

It is now possible to define the vertex operators representing
string states. We shall deal with operators in the free field
approximation. For a detailed analysis of the exact theory see
\cite{mo3,mo1}.

In general one works on $AdS_3 \times {\cal N}$ where the
vertex operators factorize as $V_{AdS_3\times {\cal N}} = V_{AdS_3}
\times V_{\cal N}$. 
In the remaining of this section we shall
consider only the $AdS_3$ part of the vertex operators. In the zero
 winding sector they may be written as
\begin{equation} \label{vertexjmm}
V_{AdS_3} = V_{j,m,\overline{m}} = \gamma^{j-m}
\overline{\gamma}^{j-\overline{m}} e^{\frac{2 j}{\alpha_+} \phi}
\end{equation}

\noindent where $j, m$ must belong to either ${\cal D}^\pm_j$ or ${\cal
C}^\alpha_{-\frac 12 + i \lambda}$ and
$m-\overline{m} \in \mathbb{Z}$ is required by
 singlevaluedness
on the spacetime coordinates $\{\gamma,\overline{\gamma}\}$. This
condition will arise more formally in the next section after introducing
the spectral flow operators \cite{mo1, supersymprod}.

The vertex operator (\ref{vertexjmm}) has the following
 OPEs with the currents
(\ref{j+bg})-(\ref{j-bg})
\begin{eqnarray}
J^+(z) V_{j,m}(z^\prime)
 &\sim& (m-j) \frac{V_{j,m+1}(z^\prime)}{z-z^\prime} \label{OPE+V}\\
J^3(z) V_{j,m}(z^\prime)
 &\sim& m \frac{V_{j,m}(z^\prime)}{z-z^\prime}
\label{OPE3V}\\
J^-(z) V_{j,m}(z^\prime ) &\sim& (m+j)
\frac{V_{j,m-1}(z^\prime)}{z- z^\prime} \label{OPE-V}
\end{eqnarray}

\noindent as required for a Kac Moody primary state.
 Excited string states can be constructed from these ones by
acting with creation modes of the currents. The conformal weight of
the operators (\ref{vertexjmm}) can be read from

\beq
T(z) V_{j,m}(z^\prime) \sim \frac{-j(j+1)}{k-2}
\frac{V_{j,m}(z^\prime)}{(z-z^\prime)^2} + \frac{\partial V_{j,m}(z^\prime
)}{z-z^\prime
} \quad .
\label{OPETV}
\eeq

Therefore any primary state in the zero winding sector can be represented
by (\ref{vertexjmm}).
How can one represent states in arbitrary winding sectors?

Two proposals can be found in the literature. One of them
relies on the bosonization of the $\beta\gamma$-system
followed by a redefinition of the scalars \cite{satoh, mizo}.
The winding number appears naturally  in this realization
after compactifying one of the light-like coordinates \cite{hhs}
(recall that $\gamma, \bar\gamma$ parametrize the boundary of $AdS_3$
which is compact in the angular direction).
The other approach implies the
factorization
$SL(2, \mathbb R) \rightarrow \frac{SL(2,\mathbb
R)}{U(1)}\times U(1)$,
as suggested by the no-ghost theorem \cite{evans}.
This proposal, that we shall follow, arises naturally in the
supersymmetric theory (see Section 3).

The strategy to introduce winding in the product theory
$\frac{SL(2,\mathbb
R)}{U(1)}\times U(1)$ is to first gauge the timelike $U(1)$ current
corresponding to the $J^3$ generator of
$SL(2, \mathbb R)$. This gives an Euclidean
theory representing a two dimensional black hole \cite{witten, dvv}. Since
one is gauging a compact $U(1)$ the winding number arises as a restriction
on the
allowed values of $m +\overline m$. However this condition disappears
when adding back a non-compact $J^3$ current. Indeed this current can be
appended in any winding sector, thus introducing a missmatch with
the gauged $U(1)$ current. This procedure allows to realize the currents
and
vertex operators in arbitrary winding sectors \cite{gepner}.

To gauge the $U(1)$ current from
$SL(2, \mathbb R)$ 
one introduces the fields
$A(z)$  and
$\overline{A}(\overline{z})$ which,  after choosing a
gauge slice, can be represented in terms of a free scalar field $X$ as
$A = -\partial X, ~ \overline{A} = -\overline{\partial} X $ \cite{dvv}
and $X(z)X(w)\sim -$ln$(z-w)$. 
Choosing a
particular gauge produces a Jacobian that can be realized by a fermionic
$bc-$system with fields
$B(z)$ and $C(z)$
having weights 1 and 0 respectively.
As usual when fixing the gauge there is a
\textit{BRST} charge which must commute with the states of the theory.
In this case one obtains
\begin{equation} \label{qbrst}
Q^{U(1)} = \oint   C(z) \, \left( J^3(z) - i \sqrt{\frac{k}{2}}
\partial X(z) \right) \, \textrm{d}z \quad .
\end{equation}

The holomorphic part of the vertex operators in this coset theory can be
naturally written as
\begin{equation} \label{vertexjmx}
V_{j,m}^{\frac{SL(2,\mathbb{R})}{U(1)}} = V_{j,m} e^{i
\sqrt{\frac{2}{k}} m X} \quad .
\end{equation}

Now we have to reintroduce the
$J^3$ current. The OPE (\ref{OPE33})
suggests the following bosonization
\begin{equation} \label{defY}
J^3(z) \equiv - i \sqrt{\frac{k}{2}} \partial Y(z)
\end{equation}

\noindent where $Y(z)$ is  a scalar field with timelike signature (recall
we are working in Euclidean $AdS_3$ and thus
$Y(z) Y(w) \sim +\ln{(z-w)}$). 

Finally the full energy momentum tensor is
\begin{equation}
T \equiv T_{\frac{SL(2,\mathbb{R})}{U(1)} \times U(1)} = \beta
\partial \gamma - \frac{1}{2} \partial \phi \partial \phi -
\frac{1}{\alpha_+}
\partial^2 \phi - \frac{1}{2} \partial X \partial X - B \partial C + \frac{1}{2}
 \partial Y \partial Y \label{Tproducto}
\end{equation}

\noindent
and the vertex operators can be written as
\begin{equation} \label{vertexjmq}
V_{j,m,p} = V_{j,m} e^{i \sqrt{\frac{2}{k}} m X} e^{i
\sqrt{\frac{2}{k}} p Y} \quad .
\end{equation}

Note that in this case there is no a priori connection between the
quantum numbers $p$ (corresponding to the $U(1)$ theory) and $m$
(corresponding to
$\frac{SL(2,\mathbb{R})}{U(1)}$). This is what allows to include the
winding number. Even though
$p$ does not depend on $m$ directly, the states represented by
(\ref{vertexjmq}) must correspond to unitary
representations of $\widehat{SL}(2,\mathbb{R})_k$. Therefore
 $p$ is restricted to $p = m
+ \frac{k}{2} w$, according to (\ref{j3tilde}) \footnote{Notice that
$m$ here is
$\widetilde{m}$ in (\ref{jmw}). We drop the tildes
 to lighten the notation.}. Therefore the final form of the vertex
operators is
\begin{equation} 
V_{j,m,w} = V_{j,m} e^{i \sqrt{\frac{2}{k}} m X} e^{i
\sqrt{\frac{2}{k}} (m + \frac{k}{2} w) Y} = \gamma^{j-m}
e^{\frac{2 j}{\alpha_+} \phi} e^{i \sqrt{\frac{2}{k}} m X} e^{i
\sqrt{\frac{2}{k}} (m + \frac{k}{2} w) Y} \quad .
\label{vertexjmw}
\end{equation}

It is easy to check that the conformal dimension of
$V_{j,m,w}$ is as expected from
(\ref{vjmw}), namely
\begin{equation} \label{pesojmw}
\Delta(V_{j,m,w}) = -\frac{j(j+1)}{k-2} - m w - \frac{k w^2}{4} \quad .
\end{equation}

Observe that the quantum numbers obtained by applying the currents
(\ref{j+bg})-(\ref{j-bg})
 to the vertex operators
(\ref{vertexjmw}) coincide with those produced by applying them to
(\ref{vertexjmm}). This indicates that (\ref{j+bg})-(\ref{j-bg})
corrrespond to the tilded currents, acting like (\ref{jmw}),
and
thus the quantum
numbers in
$V_{j,m,w}$ are actually tilded variables.
What is the correct
realization of the original currents?

It is easy to verify that the following definitions satisfy the algebra
and produce the correct quantum numbers when acting on the vertex
operators
(\ref{vertexjmw})
\begin{eqnarray}
J^+(z) &\equiv& -\beta(z) e^{i \sqrt{\frac{2}{k}} \left( X(z) + Y(z)
\right)} \quad\label{j+bg2}\\
J^-(z) &\equiv& -\left(\beta(z) \gamma^2(z) + \alpha_+ \gamma(z)
\partial \phi(z) + k \partial \gamma(z) \right)  e^{-i \sqrt{\frac{2}{k}}
\left( X(z) + Y(z) \right)} \quad \label{j-bg2} \\
J^3(z) &\equiv& -i\sqrt\frac k2 \partial Y
\end{eqnarray}

    \subsubsection{Spectral flow operators}
        \label{twist}

As mentioned above there is a formalism where the
restriction
$m-\overline{m} \in \mathbb{Z}$ appears naturally
\cite{supersymprod}.
Moreover the construction provides a method to obtain
the vertex operators in
 $w \neq 0$ sectors from those in $w = 0$. One advantage of
this
mechanism is that it allows to introduce  winding number very easily in
the
supersymmetric vertex operators and thus we review it here.

The spectral flow operator in
the theory on the product
$\frac{SL(2,\mathbb{R})}{U(1)} \times U(1)$ is defined as
\begin{eqnarray} 
{\cal F}^w (z,\overline{z}) = {\cal F}^w(z) \cdot \overline{\cal F}^w
(\overline{z})
= e^{i w \sqrt{\frac{k}{2}} \left( Y(z) +
\overline{Y}(\overline{z}) \right)} \quad .
\label{operadortwist}
\end{eqnarray}

Locality and closure of the OPEs are two important consistency
requirements.
In particular the following OPE
\begin{eqnarray}
{\cal F}^w (z,\overline{z}) V_{j,m,\overline{m},w=0}(z', \overline{z}')
&\sim& (z - z')^{-w m} (\overline{z} - \overline{z}')^{- w
\overline{m}} V_{j,m,\overline{m},w}(z', \overline{z}') \nonumber
\\
&=& (z - z')^{- (m - \overline{m}) w} | z - z' |^{-2 \overline{m}
w} V_{j,m,\overline{m},w}(z', \overline{z}') \quad\quad\quad
\label{OPEtwist}
\end{eqnarray}

\noindent
implies
that the operators
$V_{j,m,\overline{m},w}$
 must be included in the theory and $m - \overline{m} \in
\mathbb{Z}$.
It may be verified that  $V_{j,m,\overline{m},w}$
 coincide with
(\ref{vertexjmw}).

\subsection{Correlation functions and the Coulomb gas formalism}

The Coulomb gas formalism to compute correlation functions was found to be
very natural to obtain scattering amplitudes violating winding number
conservation in \cite{gn2, gn3}. Here we briefly review the basic features
of those works and in the next section we develop the supersymmetric
extension.

Correlation functions are defined as usual through an Euclidean functional
integral, namely

\begin{equation} \label{corrdef1}
\Big\langle V_{\alpha_1}(z_1) \ldots V_{\alpha_n}(z_n)
\Big\rangle_\Sigma \equiv \int \, [ \textrm{d}\phi ] \, e^{-S} \,
V_{\alpha_1}(z_1) \ldots V_{\alpha_n}(z_n)
\end{equation}

\noindent where $V_{\alpha_i}(z_i)$ are the vertex operators
(\ref{vertexjmw}) with quantum numbers $\alpha_i=j_i, m_i, \bar
m_i, w_i$, $\Sigma$ denotes the compact topology of the worldsheet
(here we shall work on the sphere) and the action $S$ is given by
(\ref{slibre3}). The measure $[ \textrm{d}\phi ]$ is a compact
notation for the measure of all fields involved. This formalism
allows to compute correlators as a perturbative expansion in the
interaction term

\begin{equation} \label{accionfull2}
 S_{int} =  \frac{1}{4\pi} \int
\textrm{d}^2 z \beta \overline{\beta} e^{-\frac{2}{\alpha_+} \phi} \quad .
\end{equation}

Scattering amplitudes are obtained from (\ref{corrdef1}) after integrating
the insertion points of the vertex operators over the complex plane
and dividing by the volume of the conformal group as
\begin{equation} \label{scatcorr}
\mathcal{A}_{\alpha_1 \ldots \alpha_n} = \frac{1}{Vol_{PSL(2,
\mathbb{C})}} \int \textrm{d}^2 z_1 \ldots \textrm{d}^2 z_n \,
\Big\langle V_{\alpha_1}(z_1) \ldots V_{\alpha_n}(z_n)
\Big\rangle_{S^2} \quad .
\end{equation}

Since the action is free except for the factor
$S_{int}$, it is also possible to define the correlators through Wick
contractions. The perturbative expansion of the functional
integral is thus reproduced by inserting powers of
$S_{int}$ into the correlators. We follow this purely algebraic procedure
(which does not rely on the action once the propagators are given)
because it provides a natural way to introduce the Coulomb gas formalism.

There are basically two types of correlators to compute: those involving
exponentials of free fields
($\phi$, $X$, $Y$) and those containing
$\beta\gamma$ fields. As mentioned above these last ones form a
$bc-$system with background
charge 1. They can be bosonized as
$\beta \cong - i \partial v e^{i v -u}~ , ~ \gamma \cong
e^{u - i v}$
where $u$ and $v$ are canonically normalized bosons with
background charge 1 and $-i$ respectively.
Therefore one only has to consider exponential operators of free fields,
eventually with a background charge $Q$ \footnote{The term $\partial v$ in
the bosonization of $\beta$ can be written as
$\beta \cong - \left( \partial e^{i v} \right) e^{-u}$, thus one can
perform Wick contractions of exponential factors in this case as well and
apply the operator
$\partial$ at the end of the calculation.}.

As usual nonvanishing correlators must satisfy the conservation law
$ Q + \sum_{i=1}^n \alpha_i
= 0 $.
This raises a problem for the two point functions of a vertex operator
with
itself which is in general expected to be nonvanishing. The solution is
provided by the conjugate vertex operators
$\widetilde{V_\alpha} = V_{-Q-\alpha}$.
The general
solution for higher point functions
was developed by
Dotsenko and Fateev in \cite{df} and the strategy is to introduce
the so called {\it screening operators} in
the correlation functions.
These insertions must not alter the conformal structure of the correlators
and therefore they must commute with the currents and have zero conformal
dimension.
These observations lead to consider the following non local operators 
\cite{zamfei, beroog}:
\beq
\mathcal{S}_+ = \int \textrm{d}^2z \, \beta(z)
\overline{\beta}(\overline{z}) e^{-\frac{2}{\alpha_+}
\phi(z,\overline{z})} \quad ; \quad
\mathcal{S}_- = \int \textrm{d}^2z \,
\beta(z)^{\frac{\alpha_+^2}{2}}
\overline{\beta}(\overline{z})^{\frac{\alpha_+^2}{2}} e^{-\alpha_+
\phi(z,\overline{z})} \quad .\label{screens}
\eeq

Consequently the correlators in string theory on $AdS_3$ can be written as
\footnote{Notice that ${\cal S}_+$ is the interaction term in the action
(\ref{accionfull2}), therefore computing the correlators
(\ref{corrbos1}) using $n_- = 0$ is completely equivalent to a
perturbative expansion of order
$n_+$ in the path integral formalism. }
\begin{equation} \label{corrbos1}
\Big\langle \mathcal{S}_+^{n_+} \, \mathcal{S}_-^{n_-} \,
V_{j_1,m_1,w_1}(z_1) \ldots V_{j_n,m_n,w_n}(z_n) \Big\rangle_{S^2}
\end{equation}

\noindent
and the conservation laws are
\begin{eqnarray}
\beta\gamma &:& \#\gamma - \#\beta + Q_{\beta\gamma} = 0 \quad
\rightarrow \quad \sum_{i=1}^n \left( j_i - m_i \right) - n_+
- \frac{\alpha_+^2}{2} n_- + 1 = 0 \quad\quad\quad\label{sl2leybg1}\\
\phi &:& \sum_{i} \alpha_i^\phi + Q_\phi = 0 \quad \rightarrow
\quad  \frac{2}{\alpha_+} \left(  \sum_{i=1}^n  j_i - n_+ -
\frac{\alpha_+^2}{2} n_- + 1 \right) = 0 \quad\label{sl2leyphi1}\\
X &:& \sum_{i} \alpha_i^X = 0 \quad \rightarrow \quad i
\sqrt{\frac{2}{k}} \sum_{i=1}^n m_i = 0 \label{sl2leyX1}\\
Y &:& \sum_{i} \alpha_i^Y = 0 \quad \rightarrow \quad i
\sqrt{\frac{2}{k}} \sum_{i=1}^n \left( m_i + \frac{k}{2} w_i
\right) = 0 \label{sl2leyY1}
\end{eqnarray}

\noindent where $\alpha_i$ represent the charge of the operators under the
various fields. Equation
(\ref{sl2leybg1}) is contained in the other three, thus they can be
summarized as

\begin{equation} \label{sl2leytot1}
\sum_{i=1}^n j_i  + 1 = n_+ + (k-2) n_- \quad ;
\quad \sum_{i=1}^n m_i = 0 \quad ; \quad \sum_{i=1}^n w_i = 0
\end{equation}

\noindent where the quantum numbers can be read from the vertex operators
$V_{j,m,w}$. In the non-compact theory
$-\frac 12 < j < \frac {k-3} 2$ for the discrete series and $j=-\frac 12
+ i \lambda$ for the principal continuous series. Therefore it is
necessary to consider the analytic extension of
(\ref{sl2leytot1}) for $n_+, n_- \in \mathbb{C}$. Actually,  once this
generalization is allowed any correlator can be computed using only one
kind of screening operators.

Similarly as in the case of minimal models it is possible to define
 conjugate operators in the $SL(2,\mathbb{R})$ WZW model.
One candidate for conjugate operator to $V_{j,m,w}$ is
$\widetilde{V_{j,m,w}} = V_{-1-j,m,w}$ \footnote{Strictly, one has
to multiply $V_{-1-j,m,w}$ by a coefficient proportional to
$\frac{\Gamma(j+1-m)}{\Gamma(-j-m)}$. In fact, this coefficient is
nothing else but the two point function. This is related to the
Fourier expansion of the square integrable functions on $AdS_3$
\cite{gks}.}. Indeed this operator has the correct conformal
dimension and OPE with the currents. One can verify also that the
two point functions $\langle V_{j_1,m_1,w_1}
V_{-1-j_2,m_2,w_2}\rangle$ do not require screening operators if
$j_2=j_1, m_2 = -m_1$ and $w_2=-w_1$ (see (\ref{sl2leytot1})). The
signs in  $m$ and $w$ refer to the distinction between {\it
ingoing} and {\it outgoing} states \footnote{One could have
included these signs in the definition of conjugation; we choose
not to do that in order to stress the conceptual idea that both
operators represent the same physical state.}.

The formalism reviewed above allows in principle to compute any
correlation function satisfying winding number conservation. However it
was suggested in \cite{fzz} and shown in
\cite{mo3} that $n$-point functions violating winding number
conservation up
to $n-2$ units can be in general nonvanishing. This was considered in
the free field approximation in \cite{gn2, gn3} where the algebraic
formulation was used to introduce new
conservation laws, thus extending the original idea designed by Dotsenko
\cite{dotse}.

To implement this procedure it is important to consider different
representations of the identity operator. The identity has zero
conformal weight and regular OPE with the currents. These
conditions are satisfied by the state $|j,m,w\rangle =
|-1,0,0\rangle$ (notice that $m=0$ singles out $j=-1$ over $j=0$
if this state is to belong to one of the discrete series
(\ref{dj+})-(\ref{dj-})). This implies that the identity is not a
physical state.

The first non trivial representation of the identity one can
consider is the conjugate operator \footnote{The free field
formalism is, in principle, valid near the boundary. This means
that, because $j=-1$, one would have to use the conjugate vertex
$\widetilde{V_{-1,0,0}}$, that dominates in the limit $\phi
\rightarrow \infty$. Indeed, this prescription gives the usual
identity operator $1$. The existence of screening operators in the
$w=0$ sectors indicates that it is also possible to obtain a
conjugate identity under $j \leftrightarrow -j-1$.} $\widetilde{1}
\equiv\mathcal{I}_{-1} = V_{-1,0,0} = \gamma^{-1}
e^{\frac{-2}{\alpha_+} \phi}$. However this realization is
obtained when conjugating with respect to the conservation laws
(\ref{sl2leytot1}) and then one cannot expect that it solves the
winding nonconservation  problem.

There is another well known representation of the identity given by
the operator $\widetilde{\mathcal{I}}_0 = \beta^{k-1}
e^{\frac{2(1-k)}{\alpha_+}\phi}$ \cite{dotse}. It leads to new
conservation laws assuring that $\Big\langle \widetilde{\mathcal{I}}_0 \, 1
\Big\rangle_{S^2}$   is non vanishing. Notice that redefining the
conjugate identity (from
$\mathcal{I}_{-1}$ to $\widetilde{\mathcal{I}}_0$) is
equivalent to redefining the out vacuum of the theory. The corresponding
conservation laws are
\begin{eqnarray}
\beta\gamma [\widetilde{\mathcal{I}}_0]&:&
\#\gamma - \#\beta = 1 - k \label{sl2leybg2}\\
\phi [\widetilde{\mathcal{I}}_0]&:&
\sum_{i} \alpha_i^\phi  = \frac{2(1-k)}{\alpha_+} \label{sl2leyphi2}\\
X [\widetilde{\mathcal{I}}_0]&:&
\sum_{i} \alpha_i^X = 0  \label{sl2leyX2}\\
Y [\widetilde{\mathcal{I}}_0]&:& \sum_{i} \alpha_i^Y = 0
\label{sl2leyY2}
\end{eqnarray}

It is interesting to note that introducing one screening operator
$\mathcal{S}_-$ in the correlation functions deduced from
$\widetilde{\mathcal{I}}_0$ one obtains the original conservation
laws (\ref{sl2leybg1})-(\ref{sl2leyY1}). This suggests that it is
possible to go from one case to the other redefining the out
vacuum through the inclusion of a screening operator
$\mathcal{S}_-$. This observation indicates that all the
correlation functions computed using
(\ref{sl2leybg2})-(\ref{sl2leyY2}), can also be computed using
(\ref{sl2leybg1})-(\ref{sl2leyY1}). Therefore this new
representation cannot solve the problem of winding
non conservation either.
There is a completely analogous statement that uses
$\mathcal{S}_+$ to relate the usual identity with
$\mathcal{I}_{-1}$.

The identities which solve the problem can be obtained recalling
the equivalence $\widehat{\mathcal{D}}_{\widetilde{j}}^{\pm , w =
\mp 1} \sim
\widehat{\mathcal{D}}_{\frac{k}{2}-2-\widetilde{j}}^{\mp , w =
0}$. Indeed, an identity allowing to violate winding conservation
must belong to one of the  $w\ne 0$ sectors, and the equivalence
between representations in $w=0$ and $w=\pm 1$ assures the
existence of such operator. Actually the replacement $ j
\rightarrow \frac{k}{2} - 2 - j$ with $j=-1$ in \footnote{Once
more the conjugate representation is used. This is the natural
thing to do for a $j=-1$ operator near the boundary.}
$\widetilde{V_{j,m,w}}$ leads to the following operators
\footnote{$m= 0$ is required by regularity of the OPE with $J^3$,
but the label in  $V_{j,m,w}$ is actually $\widetilde{m} = m -
\frac{k}{2} w$.}
\begin{eqnarray}
\mathcal{I}_+ &=& e^{-\frac{k}{\alpha_+} \phi} e^{-i
\sqrt{\frac{k}{2}} X} \label{Iplus}\\
\mathcal{I}_- &=& \gamma^{-k} e^{-\frac{k}{\alpha_+} \phi} e^{i
\sqrt{\frac{k}{2}} X} \label{Iminus} \quad .
\end{eqnarray}

\noindent They both have zero conformal weight and commute with
$J^3$. ${\cal I}_+$ (${\cal I}_-$) commutes with $ J^-$ ($ J^+)$
whereas the residue of the OPE with $J^+$ ($J^-)$ is a spurious
state which decouples in the correlators.

The conservation laws associated to
$\mathcal{I}_+$ are
\begin{eqnarray}
\beta\gamma [\mathcal{I}_+]&:& \#\gamma - \#\beta = 0 \label{sl2leybg+}\\
\phi [\mathcal{I}_+]&:& \sum_{i} \alpha_i^\phi  =
-\frac{k}{\alpha_+} \label{sl2leyphi+}\\
X [\mathcal{I}_+]&:& \sum_{i} \alpha_i^X = - i \sqrt{\frac{k}{2}}  \label{sl2leyX+}\\
Y [\mathcal{I}_+]&:& \sum_{i} \alpha_i^Y = 0 \label{sl2leyY+}
\end{eqnarray}

\noindent
whereas the laws implied by
$\mathcal{I}_-$ are
\begin{eqnarray}
\beta\gamma [\mathcal{I}_-]&:& \#\gamma - \#\beta = - k \label{sl2leybg-}\\
\phi [\mathcal{I}_-]&:& \sum_{i} \alpha_i^\phi  =
-\frac{k}{\alpha_+} \label{sl2leyphi-}\\
X [\mathcal{I}_-]&:& \sum_{i} \alpha_i^X = + i \sqrt{\frac{k}{2}}
 \label{sl2leyX-}\\
Y [\mathcal{I}_-]&:& \sum_{i} \alpha_i^Y = 0 \label{sl2leyY-}
\end{eqnarray}

It is possible to find new conjugate vertex operators with respect to
these
conservation laws.
In all these cases the operator conjugate to
 $V_{j,m,w}$ has in general a complicated form. The simplest expressions
are found
for the highest or lowest weight operators and they are given by
\footnote{Except for normalization factors that can be obtained from the
two point functions}
\begin{eqnarray}
\widetilde{V}_{j,-j-1,w} &=& \beta^{k-2j-3}
e^{\frac{2(2-k+j)}{\alpha_+} \phi} e^{i \sqrt{\frac{2}{k}} (-j-1)
X} e^{i \sqrt{\frac{2}{k}} (- j - 1 + \frac{k}{2} w) Y}
\label{vertexconj0}\\
\widetilde{V}_{j,j+1,w}^+ &=&
e^{\frac{2(j+1-\frac{k}{2})}{\alpha_+} \phi} e^{i
\sqrt{\frac{2}{k}} (j+1-\frac{k}{2}) X} e^{i \sqrt{\frac{2}{k}}
(j + 1 + \frac{k}{2} w) Y} \label{vertexconj+}\\
\widetilde{V}_{j,-j-1,w}^- &=& \gamma^{2j+2-k}
e^{\frac{2(j+1-\frac{k}{2})}{\alpha_+} \phi} e^{i
\sqrt{\frac{2}{k}} (-j-1+\frac{k}{2}) X} e^{i \sqrt{\frac{2}{k}}
(- j - 1 + \frac{k}{2} w) Y} \label{vertexconj-}
\end{eqnarray}

More general vertex operators may be constructed from these ones
applying the currents $J^\pm$. It is possible to obtain other
vertices if one makes the change \footnote{Besides possible
normalization factors} $j \leftrightarrow -j-1$ in the above
expressions. However, the choice that we made is dominant in the
$\phi \rightarrow \infty$ limit.

How do these operators solve the winding non conservation problem?
A generic correlation function in this new formalism is given by
expectation values of the form
(\ref{corrbos1}), where now the vertex operator acting on the in vacuum is
$V_{j,m,w}$ and the one acting on the out vacuum is a conjugate
operator $\widetilde V_{j,m,w}$.
This prescription specifies which realization of the identity is being
used and, consequently, which conservation laws hold. However there is no
natural choice for the intermediate vertex operators, and thus one can use
either direct ($V_{j,m,w}$) or conjugate ($\widetilde {V}_{j,m,w}$)
operators for the internal insertions. This
last possibility allows to violate winding number conservation as follows.

Notice that the conservation laws for the fields $X$ and
$Y$ associated with
$\mathcal{I}_\pm$  amount to
\footnote{Here only the conservation laws for $X$ and $Y$ are relevant
since the others can be handled through the inclusion of screening
operators.}
\begin{eqnarray}
X [\mathcal{I}_\pm]&:&  m_1 + \sum_{d.i.o.} m_i + \sum_{c.i.o.}
\left( m_i \mp \frac{k}{2} \right) + m_n \mp \frac{k}{2} = \mp
\frac{k}{2} \nonumber \\
& & \rightarrow \, \sum_{i=1}^n m_i = \pm n_c \, \frac{k}{2}
\label{conwindX1}\\
Y [\mathcal{I}_\pm]&:& \sum_{i=1}^n m_i + \frac{k}{2} \sum_{i=1}^n
w_i = 0 \nonumber\\
& & \rightarrow \, \sum_{i=1}^n w_i = \mp \, n_c \label{conwindY1}
\end{eqnarray}

\noindent where the sum over $d.i.o.$ is over \textit{direct
internal operators} and the sum over $c.i.o.$ is over the
\textit{conjugate internal operators}. $n_c$ is the number of
$c.i.o.$ in a correlation function, while $n-2-n_c$ is the number
of $d.i.o.$

Equations (\ref{conwindX1}) and (\ref{conwindY1}) explicitly exhibit
the amount of winding number non conservation of the correlators
when internal conjugate
operators are
inserted. Furthermore the maximum total winding number of a correlator is
$n-2$ since this is the maximum amount of internal operators. This result
was suggested in \cite{fzz} and demonstrated by algebraic arguments in
\cite{mo3} in the exact theory.

To finish this section let us notice that there are other
representations for the identity and conjugate operators. These
are related via screening operators to other identities/operators
in very much the same way that $\mathcal{I}_{-1}$ and 1 are. The
existence of these representations is due to an accidental
cancelation in the quadratic terms appearing in the conformal
weight that allows to consider products of identities and
screening operators. In this way one may construct yet another
conjugate identity operator using the vertices
(\ref{vertexconj0}). Indeed inserting the quantum numbers of the
identity in the sector $w=1$ one obtains
\beq \widetilde{\cal I}_+ = \beta^{-1}e^{-\frac{\alpha_+}2 \phi}
e^{-i\sqrt\frac k2 X} \quad  \eeq

\noindent This expression has to be defined through analytic
continuation since negative powers of $\beta$ cannot be understood
otherwise. The same feature can be observed in the conjugate
representations of vertices (\ref{vertexconj+}). Thus, we can
consider alternate expressions
\beq \widetilde{V}^+_{j,-j-1,w} = \beta^{-2j-2}
e^{\frac{2(j+1-\frac{k}{2})}{\alpha_+} \phi} e^{i
\sqrt{\frac{2}{k}} (-j-1-\frac{k}{2}) X} e^{i \sqrt{\frac{2}{k}}
(-j-1+\frac{k}{2} w) Y} \eeq

Once more the dominant expression in the $\phi \rightarrow \infty$
limit was chosen. This vertex was used in calculations in
\cite{gn3}.

        \subsubsection{Screening operators in $w\neq0$ sectors?}

The reader might wonder at this point whether it is possible to
incorporate winding number non-conservation in the
 Coulomb gas formalism in the usual way, $i.e.$ through screening operators in
$w\ne 0$ sectors. In that case one might avoid introducing
conjugate vertices. This is the strategy we pursued in order to
break the $j$ conservation laws. However there are several
arguments that imply this is not a possibility for winding number
violation.

We observe that in order to
violate winding conservation the hypothetical screening operators must 
be charged under $X$ or $Y$ which implies they should have an
exponential factor in $X$ or $Y$. 
It is interesting to note that
the problem of finding all
viable screening operators is dual to that of finding all possible
interaction terms for the action (\ref{slibre3}) that do not break
the original symmetries. For that reason we should require that
these operators are not only $BRST$ invariant, but that they also
have a full gauge invariant form that could be added to the action
without gauge fixing \footnote{See \cite{dvv} for an explicit
form of this action.}. This implies that we should not consider
operators charged under $Y$ for the inclusion of these in the
action would break its symmetry under $J^3$. This leads us to
consider operators charged under $X$.
However, the requirement of
gauge invariance poses another objection to the existence of these
screening operators: it is not clear how to construct an operator
with an exponential factor $X$ that could have a gauge invariant
form \footnote{Recall that the gauge field $A$ is related to the field
$X$ through derivation.}. 

Being difficult to work with full gauge invariant forms we forget
about this problem and consider candidate $BRST$ invariant
operators. Manifestly $BRST$ invariant operators that are charged
under $X$ have the form \footnote{We omit the surface integrals
for the moment. }
\begin{eqnarray}
Q_1 = \beta^{b-m} e^{i \frac{2}{\alpha_+} b \phi} e^{i
\sqrt{\frac{2}{k}} m X} \label{Q1screen}\\
Q_2 = \gamma^{m-b} e^{i \frac{2}{\alpha_+} b \phi} e^{i
\sqrt{\frac{2}{k}} m X} \label{Q2screen}
\end{eqnarray}

\noindent where $m + \frac{k}{2} w = 0$ and $b$ is fixed by the
condition of unit conformal weight. It turns out that their OPEs
with $J^\pm$ are neither regular nor give total derivatives (as in
the case of the screening operators (\ref{screens})).
This objection cannot be overcome by more
sophisticated operators either, in particular by insertions of the
form $\partial ^n X$ or $(\partial X)^n$.

Finally, we could argue that if screening operators existed in
$w\ne 0$ sectors it would be possible to violate winding number
conservation by any amount. If one considers analytic continuation
in the number of screening operators inserted in the correlators
to any real or complex value (as required by the $j$ conservation
laws) it would be possible to violate winding conservation by an
arbitrarily large non-integer number. However, as mentioned in the
previous section, Maldacena and Ooguri proved in \cite{mo3} that
winding conservation of $n$-point functions can be violated by
integer numbers bounded by $n-2$. This shows that consistency of
the free field formalism requires the non-existence of screening
operators in $w \neq 0$ sectors.

\subsection{Two and three point amplitudes}

Correlation functions in $AdS_3$ string theory have been computed
in \cite{mo3} using the exact results obtained in \cite{teschner}
for the $\frac{SL(2, \mathbb C)}{SU(2)}$ coset. The Coulomb gas
formalism was applied in \cite{becker} to calculate two and
three point functions in the free field approximation to the
$\frac{SL(2,\mathbb R)}{U(1)}$ model. The method was extended to
$SL(2,\mathbb R)$ in \cite{gn2, gn3}. The results for winding
conserving amplitudes in this approach agree with the exact ones.
Three point functions violating winding conservation were
originally computed in \cite{gn3} and these  results obtained in the
free
field approximation were later found in the exact theory in \cite{mo3}.
Here we briefly summarize a few aspects of the computations in
\cite{gn3} to facilitate the discussion of the supersymmetric case
in the next section.

It turns out that it is easier to start with three point
functions. Let us consider winding conserving amplitudes first.
The simplest correlator contains one state of highest weight in
the conjugate representation ($j \leftrightarrow -j-1$). Arbitrary
three point amplitudes can be expressed as a function of this one
acting with the lowering operator $J^-$. Indeed, applying $J^-$
one gets correlators with the insertion of one state with
$m=-1-j-N$, $N$ being the number of lowering operators. After an
analytic extension in $N$ one gets any three point function.

Fixing as usual $z_1=0, z_2=1$ and
$z_3=\infty$, the calculation factorizes into correlators of $\beta\gamma$
fields and of exponential factors.
Using screening
operators ${\cal S}_+$, the first contribution amounts to
\begin{equation} \label{bgcorrvdm}
\Big\langle \gamma^{j_2-m_2}(1) \gamma^{j_3-m_3}(\infty)
\prod_{i=1}^{n_+} \beta(y_i) \Big\rangle =
\frac{\Gamma(-j_2+m_2+n_+)}{\Gamma(-j_2+m_2)}
\prod_{i=1}^{n_+}
|1-y_i|^{-1}
\end{equation}

\noindent where $n_+= j_1 + j_2 + j_3 + 1$.

The exponential factors lead to integrals of the Dotsenko-Fateev
type \cite{df}. Putting all together, the three point amplitudes for
states in arbitrary winding sectors are
\beqa
{\cal A}_{\alpha_1, \alpha_2, \alpha_3} = \delta^2(m_1+m_2+m_3) &&\frac
{\Gamma(j_1 - m_1+1)
\Gamma(1+j_2 - m_2)
\Gamma(1+j_3 - \bar m_3)}
{\Gamma(-j_1+\bar m_1)
\Gamma(-j_2+ \bar m_2)
\Gamma(m_3 - j_3)} \times \nonumber \\
& \times & (k-2) \left [\pi\frac{\Gamma(\frac 1{k-2})}{\Gamma(1-\frac
1{k-2})}\right ]^{n_+}D(j_1, j_2,j_3)
\label{3pbos}
\eeqa

\noindent where $\alpha_i = j_i, m_i, \bar m_i, w_i$, $\sum_i w_i=0$ and
$D$ is a function
of $j_i$ which is not necessary for our purposes here (see \cite{gn3}).

The same procedure can be followed to compute correlators that do not
preserve winding number conservation. To obtain amplitudes
with $\sum_i w_i = +1(-1)$ one inserts a conjugate operator $\tilde V^+
(\tilde V^-)$ at $z_2=1$ and performs the same steps. One gets
(\ref{3pbos})
with $\delta^2(m_1+m_2+m_3) \rightarrow \delta^2(\pm \frac k2 +
m_1+m_2+m_3)$ and
$D(j_1,j_2,j_3) \rightarrow D(j_1, 1+j_2 -\frac k2, j_3)$.

The general form of the two point functions is dictated by conformal
invariance. The two insertions must have the same conformal weight
$\Delta$ (\ref{pesojmw}) and
verify the conservation laws (\ref{sl2leytot1}) if direct ($i.e.$ non
conjugate) vertices are used. This leads to the
following expression
\beqa
\Big\langle V_{j_1,m,\bar m,w}(z_1,\bar z_1) V_{j_2,-m,-\bar
m,-w}(z_2,\bar z_2)
\Big\rangle
= | z_1 & - & z_2 |^{-4 \Delta} ~ \left [ A(j_1,m,\bar m)~
\delta(j_1+j_2+1)
\right . \nonumber \\
 && \left . +~ B(j_1,m, \bar m) ~\delta(j_1-j_2) \right ]
\label{2pbos}
\eeqa

Screening operators are not necessary to compute the first term
and it is easy to see that $A(j_1,m,\bar m)=1$.

The computation of $B(j_1,m,\bar m)$
is more involved because screening operators
have to be inserted. Moreover one cannot cancel the volume of the
conformal group in (\ref{scatcorr}) since only two points can be fixed
\footnote{Actually this is also true for the first term $A(j_1,m,\bar
m)$. However we choose the normalization so that $A=1$.}. Two techniques
have been designed in \cite{becker}
to deal with this term. One of them fixes the insertion points of
the
vertex operators at $z_1=0, z_2=1$ and of one of the screening
operators at $\infty$; this cancels the full volume of the conformal
group. The other one
 considers  three point
functions in
 the limit where the additional insertion goes to the identity
($j\rightarrow i0$).
In the first case one gets
\begin{equation} \label{Bjmbecker2}
B(j,m,\bar m) = n_+ \left(-\pi\frac{\Gamma(\frac
1{k-2})}{\Gamma(1-\frac 1{k-2})}\right)^{n_+}
\frac{\Gamma(1-n_+)}{\Gamma(n_+)}
\frac {\Gamma(-\frac{ n_+}{k-2})}{\Gamma(1+\frac{n_+}{k-2})}
\frac
{\Gamma(j - m+1) \Gamma(1+j + \bar
m)}{\Gamma(-j-m)\Gamma(\bar m - j)}
\end{equation}

\noindent where
we used the $\delta(j_1-j_2)$ to define $j=j_1=j_2$ and $n_+=2j+1$.
The result obtained by the second method differs from
this one by $\frac {\delta(j_1-j_2)}{n_+}$ \footnote{This result
contains an irrelevant factor $[\pi^2 (k-2)]^{-n_+}$ with respect to
reference \cite{mo3}; notice that this factor is 1 when $n_+=0$, thus it
does not affect the term $A(j,m,\bar m)$.}.

The same outcome is produced if one uses other conservation laws with the
corresponding vertex operators.

 \section{Superstring theory on $AdS_3$}

There is a direct extension of the WZW action to the
supersymmetric case (see \cite{divecchia}) which can be
written as
\begin{equation} \label{sswzw2}
S_{SWZW} = S_{WZW} + S_{f}
\end{equation}

\noindent where $S_{WZW}$ is the bosonic WZW action and $S_{f}$
is a free fermionic action. This is a surprising result leading to the
conclusion  that the supersymmetric $WZW$ model can be decomposed into
a bosonic part and a free fermionic theory.
This interesting feature can be alternatively seen from a purely algebraic
formulation of the theory. Indeed for $SL(2,\mathbb{R})$  one can
generalize the  OPEs
(\ref{OPE+-})-(\ref{OPE33}) introducing a superfield
$\mathcal{J}^a (z,\theta) = \psi^a(z) + \theta J^a(z)$
\cite{kazama} where $\psi$ denotes the supersymmetric partner of $g$, 
an element of a bosonic representation of the group. This verifies
\begin{equation} \label{sjOPE}
\mathcal{J}^a(z_1,\theta_1) \mathcal{J}^b(z_2,\theta_2) \sim
\frac{\frac{k}{2} \eta^{ab}}{(z_1-z_2)-(\theta_1-\theta_2)} +
\frac{\theta_1-\theta_2}{z_1-z_2}\, i f^{ab}_c
\mathcal{J}^c(z_2,\theta_2)
\end{equation}

\noindent or equivalently, in components,
\begin{eqnarray}
\psi^a(z) \psi^b(w) &\sim& \frac{\frac{k}{2}}{z-w} \eta^{ab}
\label{psiOPE} \\
J^a(z) \psi^b(w) &\sim& \psi^a(z) J^b(w) \sim \frac{i f^{ab}_c
\psi^c(w)}{z-w} \label{psiJOPE} \\
J^a(z) J^b(w) &\sim& \frac{\frac{k}{2}}{(z-w)^2} g^{ab} + \frac{i
f^{ab}_c J^c(w)}{z-w} \quad . \label{JaJbOPE}
\end{eqnarray}

The theory is thus equivalent to a bosonic
\textit{Kac-Moody} algebra for $SL(2, \mathbb R)$ at level
$k$ and a fermionic
\textit{Kac-Moody} algebra of commuting currents at level $k$.
For applications to string theory it is convenient to
completely decouple both models. This is possible by defining
\begin{equation} \label{jferm}
J^a(z) = j^a(z) - \frac{i}{k} f^a_{bc} :\psi^b(z) \psi^c(z): \,
\equiv j^a(z) + j_f^a(z)
\end{equation}

\noindent where
$j(z)$ and $j_f(z)$ are bosonic currents leading to a
\textit{Kac-Moody} algebra at level
$k+2$  and a free fermionic system
respectively, for
$SL(2,\mathbb{R})$.

It is now easy to construct the energy momentum tensor and the
supersymmetry current as
\begin{eqnarray}
T(z) = \frac{1}{k} \left( j^a(z) j_a(z) - \psi^a(z) \partial
\psi_a(z) \right) \label{TSUSY} \\
T_F(z) = \frac{2}{k} \left( \psi^a(z) j_a(z) - \frac{i}{3k}
f_{abc} \psi^a(z) \psi^b(z) \psi^c(z) \right) \label{GSUSY}
\end{eqnarray}

\noindent which form a
superconformal $N=1$ theory with central charge
\begin{equation} \label{c2SUSY}
c_{SL(2,\mathbb{R})} = \frac{3}{2} + \frac{3 (k+2)}{k} \equiv
\frac{3}{2} + \frac{3 k'}{k'-2}
\end{equation}

\noindent where
$k' \equiv k+2$.

In the previous section we mentioned the possibility of
considering string propagation on $AdS_3 \times {\cal N}$. In
critical bosonic string theory the internal manifold $\mathcal{N}$
allows to modify the dimension of spacetime, but it is not
strictly necessary. However this issue is more subtle in the
supersymmetric case where spacetime supersymmetry requires an
internal theory. Actually it was shown in \cite{gepner, kazama}
that spacetime supersymmetry requires $N=2$ worldsheet
supersymmetry. In particular it was observed in \cite{kazama} that
the coset $\frac{SL(2,\mathbb{R})}{U(1)}$ possesses a natural
complex structure allowing to enhance $N=1$ to $N=2$
supersymmetry. The problem is that it is not possible to directly
extend this construction to $SL(2,\mathbb{R})$ and this is the
reason why one has to consider an internal manifold. Adding an
internal theory makes it possible to dress the $U(1)$ factor of
$\frac{SL(2,\mathbb{R})}{U(1)} \times U(1)$ with $N=2$
supersymmetry. \footnote {Several examples have been considered in the
literature. The study of \textit{NS5-branes} leads to $AdS_3
\times S^3 \times T^4$ \cite{gks, ks, monopole} which corresponds
to the $SL(2,\mathbb{R}) \times SU(2) \times U(1)^4$
supersymmetric \textit{WZW} model. The case $AdS_3 \times S^3
\times S^3 \times S^1$ \cite{s3s3s1}, which is equivalent to
$SL(2,\mathbb{R}) \times SU(2) \times SU(2) \times U(1)$, has been
reconsidered recently in the context of the $AdS_3/CFT_2$ duality
in \cite{marti} (see also \cite{lu}). These theories present an extended
$N=4$
spacetime supersymmetry.}

The general case was considered in
references \cite{rocek, berenstein} where the requirements to
achieve
spacetime supersymmetry were shown to be the following:

\begin{itemize}
  \item $\mathcal{N}$ has to be a superconformal field theory
(SCFT) with central charge

  \begin{equation} \label{cargaN}
    c_\mathcal{N} = 15 - c_{SL(2,\mathbb{R})} = \frac{21}{2} -
    \frac{6}{k}
  \end{equation}

\noindent thus ensuring total central charge $c = 15$.

  \item $\mathcal{N}$ must possess an affine
$U(1)$ symmetry. Here $\chi$ will denote the supersymmetric partner of
the $J^{U(1)}$ current.

  \item The coset theory $\frac{\mathcal{N}}{U(1)}$ must be
$N = 2$   supersymmetric. The $U(1)$ R-current of this model will
be denoted $R^{\frac{\mathcal{N}}{U(1)}}$.

\end{itemize}

A consistent spacetime supersymmetric
string theory sharing all these requisites can be built. It has at least
$N=2$
supersymmetry. In order to construct a theory with $N=1$
spacetime supersymmetry one has to take a quotient by
$Z_2$ \cite{rocek}. Furthermore it was shown in
 \cite{ari} that these conditions are not only sufficient but they are
also necessary to obtain supersymmetry.

Here we shall meet these minimal requirements using the least possible
information about
$\mathcal{N}$ so that our results will be very general.
In the following section we shall develop the basic elements of this
construction that will be necessary to obtain the vertex operators and
compute
correlation functions in this theory.

    \subsection{Spin Fields,  Supercharges and Vertex Operators}
    \label{susyoperators}

In order to construct vertex operators
it is convenient to bosonize the fermionic operators and currents.
To work with canonically normalized scalars we redefine
$\psi'^a \equiv \sqrt{\frac{2}{k}} \psi^a $
so that
\beq
\psi'^a(z) \psi'^b(w) \sim \frac{\eta^{ab}}{z-w} \quad . \label{psiOPE2}
\eeq

From now on we drop the primes to lighten the notation.

If only the three fermions in $AdS_3$ were available the bosonization
could not
be done. However the affine $U(1)$ in $\cal{N}$ makes it possible to
define  canonically
normalized bosons $H_1$, $H_2$ as \cite{rocek, friedan}
\begin{equation} \label{bosonpsichi}
\partial H_1 \equiv \psi^1 \psi^2 \quad \quad i \partial H_2
\equiv \psi^3 \chi \quad .
\end{equation}

Similarly one can bosonize the currents $J^{U(1)}$ and
$R^{\frac{\mathcal{N}}{U(1)}}$ as
\cite{supersymprod}
\begin{eqnarray}
J^{U(1)} &\equiv& i \partial W \label{bosonju} \\
R^{\frac{\mathcal{N}}{U(1)}} &\equiv& i
\sqrt{\frac{c_{\mathcal{N}/U(1)}}{3}} \partial Z = i \sqrt{3 -
\frac{2}{k}} \partial Z \label{bosonR}
\end{eqnarray}

\noindent where the scalars $W$ and $Z$ are also
canonically normalized.

The spin fields are constructed analogously to the flat case
\cite{supersymprod} as
\begin{eqnarray}
S_r^+ = e^{i r (H_1 - H_2) - \frac{i}{2} \sqrt{3 - \frac{2}{k}} Z
+ i \sqrt{\frac{1}{2 k}} W} \label{spink+} \\
S_r^- = e^{i r (H_1 + H_2) + \frac{i}{2} \sqrt{3 - \frac{2}{k}} Z
- i \sqrt{\frac{1}{2 k}} W} \label{spink-}
\end{eqnarray}

\noindent where $r = \pm \frac{1}{2}$. Notice that there are two
distinct spin fields $S^+$ and $S^-$ as required by $N=2$
spacetime supersymmetry. In fact supercharges are constructed as
usual \cite{friedan}, namely
\begin{equation} \label{susychargespin}
Q_r^\pm = (2k)^{\frac{1}{4}} \oint \textrm{d}z \,
e^{-\frac{\varphi}{2}} S_r^\pm
\end{equation}

\noindent where $\varphi$ denotes the bosonization of the ghost fields.

The presence of $W$ and $Z$ in the spin fields is unusual.
$W$ provides the supercharges (\ref{susychargespin}) with
the correct charge under the
$U(1)$ current of spacetime \textit{R-symmetry}
\cite{supersymprod}
\begin{equation} \label{Rst}
\mathcal{R} = \sqrt{2k} \oint \textrm{d}z \, J^{U(1)} \quad ,
\end{equation}

\noindent
whereas the field $Z$ carries information about the supersymmetry of
$\frac{\mathcal{N}}{U(1)}$. In fact it was shown in \cite{ari} that $Z$
can
be identified with the scalar bosonizing the current
$R^{\frac{\mathcal{N}}{U(1)}}$.

We now have all the ingredients to construct the superstring
vertex operators. Since this theory is equivalent to the product
of $SL(2, \mathbb R)_{k+2}$ times a free fermionic theory, the
vertex operators can be expected to factorize accordingly. In
particular the vertex operators of the bosonic theory will
represent states of the superstring as well, taking into account
that the quantum numbers are determined by the purely bosonic
currents $j^a$. This implies that one has to replace $k
\rightarrow k^\prime =k+2$ in the expressions of Section 2.2. In
this section we will rely, mostly, on what has been done in
\cite{supersymprod}.

Let us start by considering the states with zero winding number in
the \textit{Neveu-Schwarz} sector (NS). The most general expression for
the ground state operators is
\begin{equation} \label{sl2NSvertex}
\mathcal{V}_{j,m,q,h}^{-1} = e^{-\varphi} e^{i q W}
V_{\frac{\mathcal{N}}{U(1)}} V_{j,m}
\end{equation}

\noindent where the superindex denotes the ghost charge,
$h$ is the conformal weight of the operator in
$\frac{\mathcal{N}}{U(1)}$ and $q$ is the charge under ${\mathcal R}$.

These operators have to satisfy two physical state conditions.
First they must be on mass-shell, $i.e.$
\begin{equation} \label{pesoNS1}
\Delta\left( \mathcal{V}_{j,m,q,h}^{-1} \right) = \frac{1}{2} +
\frac{q^2}{2} + h - \frac{j (j+1)}{k} = 1 \quad .
\end{equation}

Second, they must survive the
\textit{GSO} projection or, equivalently, they must be mutually local with
the supercharges
(\ref{susychargespin}). This constraint implies

\begin{equation} \label{GSONS1}
q \sqrt{\frac{2}{k}} - q_R \in 2 \mathbb{Z} + 1
\end{equation}

\noindent where $q_R$ is the worldsheet charge of
$V_{\frac{\mathcal{N}}{U(1)}}$ under
$R^{\frac{\mathcal{N}}{U(1)}}$.
This expression exhibits the relevance of the internal theory to achieve
spacetime supersymmetry.

Let us now discuss the construction of  vertex operators at excitation
level $\frac{1}{2}$. Since
the vertices must comprise a realization of the full algebra
(with currents $J^a$ given by (\ref{jferm})) the fields
$\psi^a$ must combine with
 $V_{j,m}$ so that they belong to a representation of
$J^a$. This was done in \cite{monopole} where the following
combinations were found \footnote{There is a numerical factor,
regarding normalization, that differs from the one used in
\cite{monopole}. This difference affects correlation functions.
However, numerical factors non-depending on $j$ and $m$ are not
important for this work.}
\begin{eqnarray}
\left( \psi V_j \right)_{j+1,m} &\equiv& 2 (j+1-m)(j+1+m) \psi^3
V_{j,m} + (j+m)(j+1+m) \psi^+ V_{j,m-1} + \nonumber \\
& & (j-m)(j+1-m) \psi^- V_{j,m+1} \label{psiV+}\\
\left( \psi V_j \right)_{j,m} &\equiv& 2 m \psi^3 V_{j,m} - (j+m)
\psi^+ V_{j,m-1} + (j-m) \psi^- V_{j,m+1} \label{psiV}\\
\left( \psi V_j \right)_{j-1,m} &\equiv& 2  \psi^3 V_{j,m} -
\psi^+ V_{j,m-1} - \psi^- V_{j,m+1} \label{psiV-} \quad .
\end{eqnarray}

\noindent Here the external (internal) subindex is the eigenvalue under
$J^a$ ($j^a$). It was also shown in
\cite{monopole} that the
combination (\ref{psiV}) is not
\textit{BRST} invariant and thus there are two ways to combine the
fermionic excitations in vertex operators.
Considering for example
(\ref{psiV-}), one obtains

\begin{equation} \label{sl2NSvertex2}
\mathcal{W}_{j,m,q,h}^{-1} = e^{-\varphi} e^{i q W}
V_{\frac{\mathcal{N}}{U(1)}} \left( \psi V_j \right)_{j-1,m} \quad .
\end{equation}

The mass shell condition is now
\begin{equation} \label{pesoNS2}
\Delta\left( \mathcal{W}_{j,m,q,h}^{-1} \right) = \frac{1}{2} +
\frac{q^2}{2} + h - \frac{j (j+1)}{k} + \frac{1}{2} = 1 +
\frac{q^2}{2} + h - \frac{j (j+1)}{k} = 1 \quad ,
\end{equation}

\noindent and the \textit{GSO} projection implies

\begin{equation} \label{GSONS2}
q \sqrt{\frac{2}{k}} - q_R \in 2 \mathbb{Z} \quad .
\end{equation}

The same procedure can be applied to construct
the operator $\mathcal{X}_{j,m,q,h}^{-1}$
from the combination (\ref{psiV+}).

The construction of vertex operators in the \textit{Ramond} sector
is analogous. One now has to build representations of $J^a$ from
combinations of $S^\pm_r$ and
$V_{j,m}$. In this analysis the contributions to $S^\pm_r$ from the fields
$W$ and $Z$ are irrelevant since they can be absorbed into
redefinitions of
$e^{i q W}$ and $V_{\frac{\mathcal{N}}{U(1)}}$.
Thus one may write

\begin{equation} \label{sV-}
\left( S^\pm V_j \right)_{j-\frac{1}{2},m} \equiv
S^\pm_{\frac{1}{2}} V_{j,m-\frac{1}{2}} - S^\pm_{-\frac{1}{2}}
V_{j,m+\frac{1}{2}} \quad .
\end{equation}

The combinations possessing
$S^+$ and $S^-$ are related through conjugation. They represent states
with different charges under
$\mathcal{R}$ (\ref{Rst}).
There is of course another
representation of spin fields which couples to
$j+\frac{1}{2}$ denoted
$\left( S^\pm V_j \right)_{j+\frac{1}{2},m}$.

Following the same procedure as in the NS sector one may obtain the vertex
operator in the
\textit{picture}
$-\frac{1}{2}$, namely

\begin{equation} \label{sl2Rvertex}
\mathcal{Y}_{j,m,q,h}^{-\frac{1}{2} (\pm)} =
e^{-\frac{\varphi}{2}} e^{i q W} V_{\frac{\mathcal{N}}{U(1)}}
\left( S^\pm V_j \right)_{j-\frac{1}{2},m} \quad .
\end{equation}

\noindent The mass shell condition is

\begin{equation} \label{pesoR1}
\Delta\left( \mathcal{Y}_{j,m,q,h}^{-\frac{1}{2} (\pm)} \right) =
1 \mp \frac{q_R}{2} + \frac{q^2}{2} \pm \frac{q}{\sqrt{2 k}} + h -
\frac{j (j+1)}{k} = 1 \quad ,
\end{equation}

\noindent and the
\textit{GSO} projection implies

\begin{equation} \label{GSOR1}
q \sqrt{\frac{2}{k}} - q_R \in 2 \mathbb{Z} + 1 \quad .
\end{equation}

One can proceed similarly with
$\left( S^\pm V_j \right)_{j+\frac{1}{2},m}$.

This completes the study of supersymmetric vertex operators in the zero
winding sector for the lowest excitation levels. It is now necessary to
introduce states in nonzero winding sectors.
Similarly as in the bosonic theory one can act
on the states of the $w=0$ sector  with the spectral flow
operator.
In order to construct this operator in the supersymmetric theory one might
take the exponential of the scalar bosonizing the current $
J^3 = j^3 - i \partial H_1$.
However the operator defined as in
 (\ref{operadortwist}) is not
mutually local with the supercharges (\ref{susychargespin}). So,
one includes a \textit{twist} in the $U(1)$ factor of the model
(actually, this is a natural consequence of the complex structure
of the factor $U(1)^2$, that is required by supersymmetry
\footnote{This is the factor one obtains when formulating the
theory on $\frac{SL(2,\mathbb{R})}{U(1)} \times U(1)^2 \times
\frac{\mathcal{N}}{U(1)}$.}). Therefore the generalized spectral
flow operator is given by
\begin{equation} 
{\cal F}^w_\pm = e^{i w \sqrt{\frac{k}{2}} (\Upsilon \pm  W)}
\label{optwist2}
\end{equation}

\noindent where the field
$\Upsilon$ bosonizes $J^3$.
Notice that there are two possible spectral
flow operators, but it will turn out that only one of
them is necessary to generate all the states in the theory.

The operator (\ref{optwist2}) may be rewritten in terms of the fields
bosonizing
 $j^3$
($Y$) and the fermions ($H_1$) as follows
\begin{equation} \label{optwist3}
{\cal F}^w_\pm = e^{i w \left( \sqrt{\frac{k+2}{2}} Y + H_1 \pm
\sqrt{\frac{k}{2}} W \right)}
\end{equation}

Note that the spectral flow operation can relate states of type
$\mathcal{V}$ with others of type $\mathcal{W}$ or $\mathcal{X}$
depending on whether one acts with ${\cal F}^w_+$ or ${\cal
F}^w_-$ \cite{supersymprod}. Thus only one of them will generate
new states. In the \textit{Ramond} sector these operators generate
states related by conjugation.

Following the procedure discussed in Section 2.2.1 one can construct
vertex operators in  $w\ne 0$ sectors. These are given by
\begin{eqnarray}
\mathcal{V}_{j,m,q,h}^{-1,w} &=& e^{-\varphi} e^{i \left( q + w
\sqrt{\frac{k}{2}} \right) W} V_{\frac{\mathcal{N}}{U(1)}}
V_{j,m,w}^{susy} \label{sl2NSwvertex}\\
\mathcal{W}_{j,m,q,h}^{-1,w} &=& e^{-\varphi} e^{i \left( q - w
\sqrt{\frac{k}{2}} \right) W} V_{\frac{\mathcal{N}}{U(1)}}
\left( \psi V_{j,w}^{susy} \right)_{j-1,m} \label{sl2NSwvertex2}\\
\mathcal{Y}_{j,m,q,h}^{-\frac{1}{2},w (\pm)} &=&
e^{-\frac{\varphi}{2}} e^{i \left( q - w \sqrt{\frac{k}{2}}
\right) W} V_{\frac{\mathcal{N}}{U(1)}} \left( S^\pm V_{j,w}^{susy}
\right)_{j-\frac{1}{2},m} \label{sl2Rwvertex}
\end{eqnarray}

\noindent where we defined
\beqa \label{vjmw2}
V_{j,m,w}^{susy} & = & V_{j,m} e^{i \sqrt{\frac{2}{k'}} m X} e^{i
\sqrt{\frac{2}{k'}} (m + \frac{k'}{2} w) Y} e^{i w H_1} \nonumber \\
& = & \gamma^{j-m} e^{\frac{2 j}{\alpha_{+}'} \phi} e^{i
\sqrt{\frac{2}{k'}} m X} e^{i \sqrt{\frac{2}{k'}} (m +
\frac{k'}{2} w) Y} e^{i w H_1} \eeqa

\noindent  ($\alpha_{+}' = \sqrt{2k'-4} = \sqrt{2k}$), which has
conformal dimension

\begin{equation} \label{pesovjmw2}
\Delta\left( V_{j,m,w}^{susy} \right) = - \frac{j (j+1)}{k} - \frac{k}{4}
w^2 - m w \quad .
\end{equation}

The mass shell and \textit{GSO} projection
conditions can be obtained as usual.

This concludes the study of vertex operators in the supersymmetric theory.
We now turn to a discussion of the scattering amplitudes.

\subsection{Supersymmetric Coulomb gas formalism}

The extension of the formalism described in Section 2.3 to the
supersymmetric case is straightforward. Since the supersymmetric
model is the product of a bosonic theory and a free fermionic
theory one would expect that screening operators, identities and
conjugate vertices could be built similarly as in the bosonic
model. We will do this in a constructive manner in order to show
that there are no other possibilities given by supersymmetry that
may render the formalism inconsistent.

Let us start by discussing the screening operators. Recall that
the symmetry currents $\mathcal{J}^a$ are given by (\ref{sjOPE}),
so that requiring commutation with them  is equivalent to
demanding regular OPEs with both $J^a$ and  $\psi^a$. This implies
that the screening operators must not include fermionic fields and
consequently they will contribute to the OPE with $J^a$ only
through contractions with $j^a$ (\ref{jferm}). This is an
important result. Since the currents $j^a$ realize a level
$k^\prime$ \textit{ Kac-Moody} algebra, the screening operators in
the supersymmetric theory coincide with those in the bosonic
theory replacing $k \rightarrow k'$, namely
\begin{eqnarray}
\mathcal{S}'_+ &=& \int \textrm{d}^2z \, \beta(z)
\overline{\beta}(\overline{z}) e^{-\frac{2}{\alpha'_+}
\phi(z,\overline{z})} \label{screensusysl21}\\
\mathcal{S}'_- &=& \int \textrm{d}^2z \,
\beta(z)^{\frac{\alpha_+^{'2}}{2}}
\overline{\beta}(\overline{z})^{\frac{\alpha_+^{'2}}{2}}
e^{-\alpha'_+ \phi(z,\overline{z})} \quad . \label{screensusysl22}
\end{eqnarray}

It is easy to verify that these operators have zero conformal
dimension.

Therefore the conservation laws can be modified in the same way it
was discussed for the bosonic case. Let us stress that these
screening operators do not allow to alter the conservation laws
either of the fermionic fields or of the fields in the internal
theory.

What about the identity operators?

One can proceed as in the bosonic case, applying the spectral flow
operator on the identity and then replacing $j \rightarrow
\frac{k}{2} - 2 - j$. Since the identity is not in the physical
spectrum  it will not necessarily be written as one of the vertex
operators described in the previous section. Actually the operator
$1$ is still of the form
 $V_{j,m}$ with $m=0$ and $j=-1$ as in the bosonic case.
This raises several observations. First, as discussed for the
screening operators, the condition of regularity of the OPE with
the currents implies that the identity operator must not contain
fermionic fields. Second, this condition must hold for the
symmetry currents of the full spacetime, that is $AdS_3 \times
U(1) \times \frac{\mathcal{N}}{U(1)}$. Thus the identity cannot be
charged under the field $W$ or depend on $\chi$. Finally, the
natural form of the identity $1$ is
 in the picture 0. These comments indicate that a good
starting point to find new representations of the identity operator is
 the spectral flow of $V_{j,m}$.
 This is as in the bosonic case except that the spectral flow
operator (\ref{optwist2}) now has contributions from the $U(1)$
factor as well as from the fermions.
Therefore the general form of the candidate conjugate identity operator
is

\begin{equation} \label{idconjsusy}
\mathcal{I}_* = e^{i a H_1} e^{i (q \pm \sqrt{\frac{k}{2}} w) W}
V_{j,m,w}^{susy}
\end{equation}

\noindent with $w=\pm 1$. Notice that the zero modes of $H_1$ and
$W$ are shifted in the same way that the field $Y$ was shifted for
the bosonic case. This is
 interpreted
as the string winding around
$AdS_3 \times U(1)$. Consequently the quantum numbers $a$, $m$ and $q$
are actually  tilded variables (similarly as $m$ in the bosonic case).
But once
again we omit the tildes in the operators.

Regular OPEs with the fermions $\psi^\pm$ imply \footnote{Here we
use the explicit expression (\ref{vjmw2}) for $V_{j,m,w}^{susy}$
.} $a + w = 0$, while the OPEs with the $U(1)$ symmetry currents
 $\chi$ and $\partial
W$ determine that there cannot be fermionic contributions from $U(1)$
and $q \pm
\sqrt{\frac{k}{2}} w = 0$ respectively.
These constraints lead to the following expression

\begin{equation} \label{idconjsusy2}
\mathcal{I}_* = e^{- i w H_1} V_{j,m,w}^{susy} = \gamma^{j-m}
e^{\frac{2 j}{\alpha_{+}'} \phi} e^{i \sqrt{\frac{2}{k'}} m X}
e^{i \sqrt{\frac{2}{k'}} (m + \frac{k'}{2} w) Y} \quad .
\end{equation}

Finally, plugging in the quantum numbers of the identity $m +
\frac{k'}{2} w = 0$ and $j = \frac{k'}{2} -1$, one obtains the
same operators as in the bosonic theory with the replacement $k
\rightarrow k'$, namely
\begin{eqnarray}
\mathcal{I}'_+ &=& e^{-\frac{k'}{\alpha'_+} \phi} e^{-i
\sqrt{\frac{k'}{2}} X} \label{Iplus2}\\
\mathcal{I}'_- &=& \gamma^{-k'} e^{-\frac{k'}{\alpha'_+} \phi}
e^{i \sqrt{\frac{k'}{2}} X} \quad .\label{Iminus2}
\end{eqnarray}

The conservation laws dictated by these operators coincide with the
bosonic ones except for the change
$k \rightarrow k'$. The same conclusion holds for the conjugate identities
$\mathcal{I}_{-1}$, $\widetilde{\mathcal{I}}_0$ and
$\tilde{\mathcal{I}}_+$.

Analogously to the bosonic case one can write conjugate vertex
operators with respect to these identities. One has only to
replace the factor $V_{j,m,w}$ by its conjugate versions
(\ref{vertexconj0})-(\ref{vertexconj-}), thus obtaining

\begin{eqnarray}
\widetilde{\mathcal{V}}_{j,m,q,h}^{* -1,w} &=& e^{-\varphi} e^{i
\left( q + w \sqrt{\frac{k}{2}} \right) W}
V_{\frac{\mathcal{N}}{U(1)}} e^{i w H_1}
\widetilde{V'_{j,m,w}}^* \label{sl2NSwvertexc}\\
\widetilde{\mathcal{W}}_{j,m,q,h}^{* -1,w} &=& e^{-\varphi} e^{i
\left( q - w \sqrt{\frac{k}{2}} \right) W}
V_{\frac{\mathcal{N}}{U(1)}}
\left( \psi e^{i w H_1} \widetilde{V'_{j,w}}^* \right)_{j-1,m} \label{sl2NSwvertex2c}\\
\widetilde{\mathcal{Y}}_{j,m,q,h}^{* -\frac{1}{2},w (\pm)} &=&
e^{-\frac{\varphi}{2}} e^{i \left( q - w \sqrt{\frac{k}{2}}
\right) W} V_{\frac{\mathcal{N}}{U(1)}} \left( S^\pm e^{i w H_1}
\widetilde{V'_{j,w}}^* \right)_{j-\frac{1}{2},m}
\label{sl2Rwvertexc}
\end{eqnarray}

\noindent where $\widetilde{V'_{j,m,w}}^*$ refers to any of
(\ref{vertexconj0})-(\ref{vertexconj-})\footnote{At least for
highest weight operators; the others can be obtained applying
lowering modes of the currents.} with $k \rightarrow k'$.

To finish this section let us notice that conjugation involves
only the bosonic part of $AdS_3$. Therefore the associated
conservation laws cannot break the conservation of the quantum
number $q$ or the fermionic excitations. This fact has important
consequences that are not observed in the bosonic case. In
particular, correlation functions for states with different
excitation numbers will exhibit a different winding number
violation pattern. While some choice of winding violation is
nonvanishing for one case, it will vanish identically for other
cases, involving excited states. We shall ellaborate on this point
in the following sections.

    \subsection{Supersymmetric Correlators: the Neveu-Schwarz sector}
    \label{corNS}

In this section we compute correlation functions in the
Neveu-Schwarz sector, both satisfying and violating winding number
conservation.

        \subsubsection{Two point functions in the Neveu-Schwarz sector}

The two point functions can be computed by either one of the two methods
discussed in Section 2.3.2. Since the insertion that one should consider
in
order
to obtain the result as the limit of a three point function  is purely
bosonic, both methods turn out to be identical.

Recall that the ghost charge of the correlators must be $-2$.
Therefore we can use the natural form of the vertex operators in the
picture $-1$. Moreover the direct version of the vertices can be taken
since violation of winding number conservation is not expected.
There are thus three different types of two point functions one
can consider; those with: two insertions of
$\mathcal{V}$, two insertions of
$\mathcal{W}$ and  one insertion of each kind.

We shall omit the contributions from the internal
$\frac{\mathcal{N}}{U(1)}$ theory and thus the computations correspond to
an $AdS_3 \times U(1)$ background.

Let us start by considering the correlation functions of two ground states
 $\mathcal{V}$:
\begin{eqnarray}
\Big\langle \mathcal{V}_{j_1,m_1,q_1}^{-1,w_1}(z_1) \,
\mathcal{V}_{j_2,m_2,q_2}^{-1,w_2}(z_2) \Big\rangle &=&
\Big\langle e^{-\varphi(z_1)} e^{-\varphi(z_2)} \Big\rangle
\Big\langle e^{i \left(q_1 + w_1 \sqrt{\frac{k}{2}}\right) W(z_1)}
e^{i \left(q_2 + w_2 \sqrt{\frac{k}{2}}\right) W(z_2)} \Big\rangle
\nonumber\\
& & \Big\langle e^{i w_1 H_1(z_1)} e^{i w_2 H_2(z_2)} \Big\rangle
\Big\langle V'_{j_1,m_1,w_1}(z_1) V'_{j_2,m_2,w_2}(z_2)
\Big\rangle \quad \label{cNSVV}
\end{eqnarray}

\noindent
Note that this
factorization is possible since the screening operators in the
supersymmetric theory coincide with those in the bosonic model.
Fixing as usual
$z_1=0$ and
$z_2=1$, we obtain
\begin{eqnarray}
\Big\langle \mathcal{V}_{j_1,m_1,q_1}^{-1,w_1}(0) \,
\mathcal{V}_{j_2,m_2,q_2}^{-1,w_2}(1) \Big\rangle=
\Big\langle V'_{j_1,m_1,w_1}(0) V'_{j_2,m_2,w_2}(1) \Big\rangle
\label{cNSVV2}
\end{eqnarray}

\noindent
where the conservation laws for $W$ and $H_1$ are the following
\begin{eqnarray}
W&:& \quad q_1 + q_2 + \sqrt{\frac{k}{2}} \left( w_1 + w_2 \right)
=
0 \label{cNSVVleyW}\\
H_1&:& \quad  w_1 + w_2 = 0 \label{cNSVVleyH1}
\end{eqnarray}

The two point functions (\ref{cNSVV2}) were computed in the free field
approximation
in \cite{gn3} and the result (\ref{2pbos}) is quoted in Section 2.
Here we only have to change $k \rightarrow k'$.

Let us now consider  two point functions of type
$\langle \mathcal{W} \mathcal{W} \rangle$, namely
\begin{eqnarray}
\Big\langle \mathcal{W}_{j_1,m_1,q_1}^{-1,w_1}(z_1) \,
\mathcal{W}_{j_2,m_2,q_2}^{-1,w_2}(z_2) \Big\rangle &=&
\Big\langle e^{-\varphi(z_1)} e^{-\varphi(z_2)} \Big\rangle
\Big\langle e^{i \left(q_1 - w_1 \sqrt{\frac{k}{2}}\right) W(z_1)}
e^{i \left(q_2 - w_2 \sqrt{\frac{k}{2}}\right) W(z_2)} \Big\rangle
\nonumber\\
& & \Big\langle \left(\psi V^{susy}_{j_1,w}\right)_{j_1-1,m_1}(z_1)
\left(\psi V^{susy}_{j_2,w}\right)_{j_2-1,m_2}(z_2) \Big\rangle
\label{cNSWW}
\end{eqnarray}

Here the contribution from the field $W$ leads to
\begin{equation}
W: \quad q_1 + q_2 - \sqrt{\frac{k}{2}} \left( w_1 + w_2 \right) =
0  \quad . \label{cNSWWleyW}
\end{equation}

In order to deal with the terms
$\left(\psi V \right)$
it is convenient to bosonize the fermions as
$\psi^+ = \sqrt 2 e^{i H_1} ;
\psi^- = {\sqrt{2}} e^{-i H_1}$
 which leads to
\begin{eqnarray}
T^+ &\equiv& \psi^+ e^{i w H_1} V'_{j,m-1,w} = \sqrt{2} e^{i (w+1)
H_1} V'_{j,m-1,w}
\label{boson-3+}\\
T^- &\equiv& \psi^- e^{i w H_1} V'_{j,m+1,w} = \sqrt{2} e^{i (w-1)
H_1} V'_{j,m+1,w} \label{boson-3-}
\end{eqnarray}

Notice that this implies the conservation laws $w_1 + w_2 \pm 2 =
0$ from the field $H_1$ for the terms $\langle T^\pm T^\pm
\rangle$, whereas the factors $V'_{j,m,w}$ give $w_1+w_2 = 0$.
Therefore the only nonvanishing contributions arise from
contractions of $\psi^3$ and from terms containing $\langle T^\pm
T^\mp \rangle$. The commutation relations $ [ J^\pm_0, V_{j,m} ] =
(m \mp j) V_{j,m\pm1} $ can be used to replace the vertex
operators $V_{j,m+1}$ or $V_{j,m-1}$ and after a little algebra
one obtains
\begin{eqnarray}
\Big\langle \mathcal{W}_{j_1,m_1,q_1}^{-1,w_1} \,
\mathcal{W}_{j_2,m_2,q_2}^{-1,w_2} \Big\rangle &=&
\frac{4j(2j+1)}{m^2-j^2} \Big\langle V'_{j_1,m_1,w_1}
V'_{j_2,m_2,w_2} \Big\rangle \label{cNSWW4}
\end{eqnarray}

\noindent for
$j_1=j_2 = j$ and
$m_1=-m_2
\equiv m$ \footnote{The same result was found in \cite{daniel}.}.

Lastly let us discuss the mixed correlator
$\langle \mathcal{V}
\mathcal{W} \rangle$. It is easy to see that this vanishes identically.
Indeed the term with $\psi^3$ leaves an unpaired fermion whereas the
conservation laws from the field $H_1$ and the operators
$V_{j,m,w}$ are incompatible.

This concludes the analysis of the two point functions in the NS sector.
In
the next section we compute three point functions.

        \subsubsection{Three point functions in the Neveu-Schwarz sector}

Here there are additional complications to take into account.
Firstly the ghost charge asymmetry implies that one has to consider one
operator in the picture 0.
Secondly the three point functions can in principle violate winding number
conservation and thus one has to consider different conservation laws.

In order to obtain the vertex operator in the picture 0 we have to
construct the picture changing operator. Generalizing the
standard procedure followed in the flat theory, where
the picture changing operator is
$P_{+1} = e^\varphi T_F$, we find for string theory in
$AdS_3$
\begin{equation} \label{sopchpicads3}
P_{+1}^{AdS_3} = \frac{e^\varphi}{\sqrt{2k}} \left\{ - 2 \psi^3
j^3 + \psi^+ j^- + \psi^- j^+ + \psi^- \psi^+ \psi^3 \right\}
\end{equation}

\noindent where the
supercorrent $T_F$ was defined in
(\ref{GSUSY}).

To obtain the picture changing operator in the full theory
one has to add
$T_F^{U(1)}$ and
$T_F^{\frac{\mathcal{N}}{U(1)}}$. The first term is easy
to write since the $U(1)$ WZW model is a flat theory, thus
\begin{equation} \label{ssuperGU1}
T_F^{U(1)} = \chi \partial W \quad .
\end{equation}

In order to define  $T_F^{\frac{\mathcal{N}}{U(1)}}$
we have to choose a particular
internal theory. However this is not necessary for the computation
of the three point functions of ground states in the NS sector since they
do not contain fermionic contributions (either from
 $\chi$ or from any fermion in
$\frac{\mathcal{N}}{U(1)}$) and thus their contraction with
$T_F$ gives one fermion in the picture 0 vertex operator that cannot be
paired.
Therefore we can define

\begin{equation} \label{sopVpic0}
\mathcal{V}^0(z) = \lim_{w \rightarrow z} P_{+1}^{AdS_3}(w)
\mathcal{V}^{-1}(z)
\end{equation}

\noindent and thus
\beqa
\mathcal{V}^{0,w}_{j,m,q} = \frac{e^{i \left( q +
\sqrt{\frac{k}{2}} w \right) W}}{\alpha'_+} \left\{ - 2 m \psi^3
e^{i w H_1}
V'_{j,m,w}
   +  \sqrt{2} (m-j) e^{i (w-1) H_1} V'_{j,m+1,w} \right. \nonumber\\
 \left.  +  \sqrt{2} (m+j) e^{i (w+1) H_1} V'_{j,m-1,w}
\right\}\label{sopVpic02} \eeqa \noindent Similarly for ${\cal W}$
we obtain \footnote{In this case we consider states in the $w=0$
sector. General states can be obtained in a similar way
considering contractions from the fermionic part of the spectral
flow operator. We leave the details to the reader.}
\begin{eqnarray}
\mathcal{W}_{j,m,q}^{0,w=0} = \sqrt{\frac{2}{k}} e^{i q W} \left\{
k' \partial \gamma V'_{j,m+1,w=0} - j \psi^- \psi^3 V'_{j,m+1,w=0}
\right. \nonumber
\\
\left. - j \psi^3 \psi^+ V'_{j,m-1,w=0} - j \psi^+ \psi^-
V'_{j,m,w=0} \right\} \quad .\label{sopWpic02}
\end{eqnarray}

We can now compute the three point functions. Let us start with
correlators of the type $\langle \mathcal{V} \mathcal{V}
\mathcal{V} \rangle$. Since the vertex operators
$\mathcal{V}^{-1}$ do not contain $\psi^3$ the first term in
(\ref{sopVpic02}) will not contribute. There are thus only two
terms to consider, namely \footnote{In the following expressions a
$\frac{\sqrt{2}}{\alpha'_+}$ factor appears. This is related to the
supercurrent and it can be absorbed
after properly normalizing the picture 0 vertex operators. We
disregard this normalization for it is irrelevant for our
purposes.}
\begin{eqnarray}
\Big\langle \mathcal{V}_{j_1,m_1,q_1}^{-1,w_1} \,
\mathcal{V}_{j_2,m_2,q_2}^{0,w_2} \,
\mathcal{V}_{j_3,m_3,q_3}^{-1,w_3} \Big\rangle =
\frac{\sqrt{2}}{\alpha'_+} \Big\langle e^{i \left( q_1 +
\sqrt{\frac{k}{2}} w_1 \right) W} e^{i \left( q_2 +
\sqrt{\frac{k}{2}} w_2 \right) W} e^{i \left( q_3 +
\sqrt{\frac{k}{2}} w_3 \right) W} \Big\rangle \nonumber\\
\left\{ (m_2-j_2) \Big\langle e^{i w_1 H_1} e^{i \left( w_2 - 1
\right) H_1} e^{i w_3 H_1} \Big\rangle \Big\langle
V'_{j_1,m_1,w_1} V'_{j_2,m_2+1,w_2} V'_{j_3,m_3,w_3}
\Big\rangle\right. \nonumber\\
\left. + (m_2+j_2) \Big\langle e^{i w_1 H_1} e^{i \left( w_2 + 1
\right) H_1} e^{i w_3 H_1} \Big\rangle \Big\langle
V'_{j_1,m_1,w_1} V'_{j_2,m_2-1,w_2} V'_{j_3,m_3,w_3} \Big\rangle
\right\}\quad \label{cNSVVV1}
\end{eqnarray}

\noindent where we fix as usual $z_1 = \infty$, $z_2 = 0$ and $z_3 =
1$. The conservation law associated to the field
 $W$ is the following for both terms
\begin{equation} \label{leyWcNSVVV}
q_1 + q_2 + q_3 + \sqrt{\frac{k}{2}} \left( w_1 + w_2 + w_3
\right) = 0
\end{equation}

\noindent whereas the field $H_1$ gives
\begin{equation} \label{leyHcNSVVV}
w_1 + w_2 + w_3 \mp 1 = 0
\end{equation}

\noindent with the $-$ ($+$) sign corresponding to the first (second) term
in
(\ref{cNSVVV1}). This is a very interesting result. In fact notice that
the correlator
$\langle
\mathcal{V} \mathcal{V} \mathcal{V} \rangle$ is nonvanishing only if
the winding number is not conserved.
Moreover the expressions
(\ref{leyWcNSVVV}) and
(\ref{leyHcNSVVV}) show that non trivial contributions from
the factor $U(1)$ other than those arising from the spectral flow
sector are necessary to obtain nonvanishing correlators.

Therefore the explicit computation of
(\ref{cNSVVV1}) requires a conjugate operator in the internal
position either according to
$\mathcal{I}_+$ or to $\mathcal{I}_-$ and only one term
will contribute in each of these cases.
For example using
$\mathcal{I}_-$ one obtains
\begin{eqnarray}
\Big\langle \mathcal{V}_{j_1,m_1,q_1}^{-1,w_1} \,
\mathcal{V}_{j_2,m_2,q_2}^{0,w_2} \,
\mathcal{V}_{j_3,m_3,q_3}^{-1,w_3} \Big\rangle =
\frac{\sqrt{2}}{\alpha'_+} (m_2-j_2) \Big\langle V'_{j_1,m_1,w_1}
\widetilde{V'}^-_{j_2,m_2+1,w_2} V'_{j_3,m_3,w_3} \Big\rangle
\quad \label{cNSVVV2}
\end{eqnarray}

\noindent with the following conservation laws
\begin{eqnarray}
m_1 + m_2 + m_3 + 1 + \frac{k}{2} = 0 \label{leymcNSVVV}\\
w_1 + w_2 + w_3 - 1 = 0 \label{leywcNSVVV}
\end{eqnarray}

Notice that these conditions are compatible with (\ref{leyHcNSVVV}).
The same procedure can be followed for $\mathcal{I}_+$. In this case
winding conservation will be
violated by one unit with the opposite sign and the other term will
survive in
(\ref{cNSVVV1}).

Using the explicit form of the bosonic correlator (\ref{3pbos}) it is easy
to
rewrite (\ref{cNSVVV2}) as
\begin{eqnarray}
\Big\langle \mathcal{V}_{j_1,m_1,q_1}^{-1,w_1} \,
\mathcal{V}_{j_2,m_2,q_2}^{0,w_2} \,
\mathcal{V}_{j_3,m_3,q_3}^{-1,w_3} \Big\rangle =
\frac{\sqrt{2}}{\alpha'_+} \Big\langle V'_{j_1,m_1,w_1}
\widetilde{V'}^-_{j_2,m_2,w_2} V'_{j_3,m_3,w_3} \Big\rangle \quad
\label{cNSVVV3}
\end{eqnarray}

\noindent and then
the amplitudes of type
$\langle \mathcal{V} \mathcal{V}
\mathcal{V} \rangle$
coincide with the bosonic ones.

Here it is interesting to comment on an observation by Giveon and
Kutasov \cite{litstring}. They suggested that correlators of
operators $\mathcal{V}$ should present a natural framework to
violate winding conservation in the coset model
$\frac{SL(2,\mathbb{R})}{U(1)}$ . This feature is related in their
work to the picture changing operator, $i.e.$ the $-\frac{1}{2}$
mode of the  supercurrent $T_F$. Actually, because of the $N=2$
supersymmetry, one can decompose $G_{-\frac{1}{2}}$ into
eigenstates under the \textit{R-symmetry} current. Then one can
choose the term with either positive or negative charge. This
gives two possible picture changing operators each one affecting
the $m$ conservation law in $\pm \frac{k}{2}$ and from this one
can read the violation to winding conservation. We have presented
here the first explicit calculation of this fact in the free field
formalism. It is also interesting to notice that in the formalism
we have developed one can read the two options proposed by Giveon
and Kutasov for the picture changing operator in (\ref{cNSVVV1})
and observe that only one of them contributes to a given
correlator. In order for any of these two correlation functions to
be non-zero it is necessary to guarantee that the bosonic part of
the correlator will have a set of conservation laws that match
those that are obtained in the fermionic part. This is possible,
in the free field formalism, only because of the existence of the
conjugate identities in the $ w = \pm 1$ winding sectors. This is
true not only for the coset but for the full $SL(2,\mathbb{R})$
$WZW$ model.

This concludes the computation of the three point functions of  type
$\langle \mathcal{V} \mathcal{V} \mathcal{V} \rangle$. The same
procedure can be extended to the other three point functions. For
example $\langle \mathcal{V} \mathcal{W} \mathcal{V} \rangle$ and
$\langle \mathcal{V} \mathcal{W} \mathcal{W} \rangle$. The
structure of the winding violation pattern is more complicated as
one adds excited fields of type $\mathcal{W}$. However, it is easy
to see that the result is zero in the last case unless $w$
conservation is violated, analogously to the case of $\langle
\mathcal{V} \mathcal{V} \mathcal{V} \rangle$. On the other hand
the function $\langle \mathcal{V} \mathcal{W} \mathcal{V} \rangle$
presents all the possibilities regarding $w$ conservation.

Considering amplitudes of type $\langle
\mathcal{W} \mathcal{W} \mathcal{W} \rangle$  introduces a
 difficulty related to the factor $\partial\gamma$ in ${\cal W}^0$.
However the formalism developed in \cite{becker} to compute
$\beta\gamma$ correlators can be easily adapted to deal with this
case. It is interesting, though, that in this particular case the
supersymmetric correlation functions are not proportional to the
bosonic ones involving only ground states, but have factors proportional
to derivatives. This
is analogous to what happens in flat space.

    \subsection{Supersymmetric Correlators: the Ramond sector}

In this section we discuss correlation functions involving
states in the Ramond sector. These have not been considered previously
in the literature. We compute amplitudes in arbitrary winding sectors and
consider the possibility of violating this quantum number.

        \subsubsection{Two point functions in the Ramond sector}

In order to achieve ghost charge $-2$ in the two point functions
in the Ramond sector we need 
vertex operators in the {\it picture} $-\frac 32$. The non trivial BRST
invariant vertices are \footnote{These can be obtained applying
considerations discussed in
\cite{rocek} and \cite{givpak}.}

\begin{equation} 
\mathcal{Y}_{j,m,q,h}^{-\frac{3}{2} (\pm)} =
e^{-\frac{3 \varphi}{2}} e^{i q W} V_{\frac{\mathcal{N}}{U(1)}}
\left( S^{-\frac 32,\pm} V_j \right)_{j-\frac{1}{2},m} 
\end{equation}

\noindent where
\beqa
 S^{-\frac 32,\pm}_r = e^{ir(H_1\pm H_2)\mp \frac i2\sqrt{3-\frac
2k}Z\pm i\sqrt{\frac1{2k}}W} \quad .
\eeqa

Therefore we have to compute the correlator
\beqa 
\Big\langle \mathcal{Y}_{j_1,m_1,q_1}^{-\frac 32, w_1, (\pm)}(z_1) &&
\mathcal{Y}_{j_2,m_2,q_2}^{-\frac 12, w_2, (\pm)}(z_2) \Big\rangle =
\Big\langle e^{-\frac{3\varphi(z_1)}{2}}
e^{-\frac{\varphi(z_2)}{2}} \Big\rangle \times \nonumber \\
&&
\Big\langle
e^{i\left(q_1 - \sqrt{\frac{k}{2}} w_1 \right) W} 
\left( S^{-\frac 32, \pm} V_{j_1}^{w_1} \right)_{j_1-\frac{1}{2},
m_1}(z_1)
e^{i\left(q_2 - \sqrt{\frac{k}{2}} w_2 \right) W}
 \left(S^\pm V_{j_2}^{w_2} \right)_{j_2-\frac{1}{2},
m_2}(z_2) \Big\rangle \nonumber
\end{eqnarray}

The conservation law for the field $H_2$ implies that terms containing
spin fields of the same (opposite) type $(\pm)$, namely $ S^{-\frac
32,\pm}_r$ and
 $S^{\pm}_{r'}$, 
vanish if $r\ne r'$ ($r=r'$). Similarly the conservation law from $H_1$
implies that correlators 
contaning spin fields of the same type 
 must violate winding number
in one unit if $r=r'$. 
 Therefore two point functions involving operators of the same type vanish
since winding number must be conserved. We stress that the $H_2$
field is responsible for inhibiting this channel.
We thus compute the nontrivial correlator, where winding conservation
equals $H_2$ conservation,
\begin{eqnarray}
\Big\langle \mathcal{Y}_{j_1,m_1,q_1}^{-\frac{3}{2},w_1, (\pm)} \,
\mathcal{Y}_{j_2,m_2,q_2}^{-\frac{1}{2},w_2, (\mp)}
\Big\rangle = - \Big\langle e^{i\left(q_1 - \sqrt{\frac{k}{2}}
w_1 \pm \frac{1}{\alpha'_+} \right) W} 
e^{i\left(q_2 -
\sqrt{\frac{k}{2}} w_3 \mp \frac{1}{\alpha'_+} \right) W}
\Big\rangle \quad \nonumber\\
\left\{ \Big\langle e^{\pm\frac{i}{2} H_2} e^{\mp\frac{i}{2} H_2}
\Big\rangle \Big\langle e^{i \left( + \frac{1}{2} + w_1 \right)
H_1} e^{i \left( - \frac{1}{2} + w_2 \right) H_1}
\Big\rangle \Big\langle V'_{j_1,m_1-\frac{1}{2},w_1}
V'_{j_2,m_2+\frac{1}{2},w_2} \Big\rangle
\right.\nonumber\\
\left. +\Big\langle e^{\mp\frac{i}{2} H_2} e^{\pm\frac{i}{2} H_2}
\Big\rangle \Big\langle e^{i \left( - \frac{1}{2} + w_1 \right)
H_1} e^{i \left( + \frac{1}{2} + w_2 \right) H_1}
\Big\rangle \Big\langle V'_{j_1,m_1+\frac{1}{2},w_1}
 V'_{j_2,m_2-\frac{1}{2},w_2} \Big\rangle
\right\}\nonumber
\end{eqnarray}

\noindent
with the following conservation laws

\begin{eqnarray}
& &q_1 + q_2  - \sqrt{\frac{k}{2}} \left(w_1 + w_2 \right)
 = 0 \\
& &w_1 + w_2 = 0 \quad . 
\end{eqnarray}

We can again use the commutator $[J^+_0 V_{j,m}]$ in the last factor
obtaining
\beq
\Big\langle \mathcal{Y}_{j_1,m_1,q_1}^{-\frac{3}{2},w_1, (\pm)} \,
\mathcal{Y}_{j_2,m_2,q_2}^{-\frac{1}{2},w_2, (\mp)}
\Big\rangle = \frac {2m}{j+\frac 12 -m} 
\Big\langle V'_{j_1,m_1-\frac{1}{2},w_1}
V'_{j_2,m_2+\frac{1}{2},w_2} \Big\rangle
\eeq
\noindent where $j=j_1=j_2$ and $m=m_1=-m_2$.

There is an alternative way to perform this computation. Mimicking
the bosonic calculation
one can take
the limit of a three point function containing an identity
operator. However inserting an operator of the type
$\mathcal{V}_{j,m}$ and taking the limit
$j\rightarrow -1$, $m \rightarrow 0$ produces a factor
$e^{-\varphi}$ which is clearly not the identity (actually this operator
has conformal dimension
$\frac{1}{2}$). On the other hand, an operator of the type
$V_{j,m}$ gives the identity when taking the appropriate limit but it does
not satisfy the
ghost charge condition.
We propose to proceed as follows. There is a representation of the
identity in the supersymmetric theory which is very useful, namely
\begin{equation} \label{isusy}
\mathcal{I}^{-2}(z) = e^{-2\varphi(z)}
\end{equation}

So we can apply the picture
changing operator to one of the R vertices and compute
\begin{equation} 
\Big\langle \mathcal{Y}_{j_1,m_1,q_1}^{w_1, (\pm)} \,
\mathcal{Y}_{j_2,m_2,q_2}^{w_2, (\pm)} \Big\rangle =
\Big\langle \mathcal{Y}_{j_1,m_1,q_1}^{-\frac{1}{2},w_1, (\pm)} \,
e^{-2\varphi} \, \mathcal{Y}_{j_2,m_2,q_2}^{\frac{1}{2},w_2,
(\pm)} \Big\rangle \quad .
\end{equation}

We now move on to three point functions.

        \subsubsection{Three point functions in the Ramond sector}

The simplest three point function one may consider is of the form
 $\langle \mathcal{Y} \mathcal{V} \mathcal{Y}
\rangle$. Notice that in this case the charge asymmetry condition can be
satisfied when all the vertex operators take their natural form.
Therefore the correlation function is as follows

\begin{eqnarray}
\Big\langle \mathcal{Y}_{j_1,m_1,q_1}^{-\frac{1}{2},w_1, (\pm)} \,
\mathcal{V}_{j_2,m_2,q_2}^{-1,w_2} \,
\mathcal{Y}_{j_3,m_3,q_3}^{-\frac{1}{2},w_3, (\pm)}
\Big\rangle = 
\Big\langle e^{-\frac{\varphi(z_1)}{2}}
e^{-\varphi(z_2)} e^{-\frac{\varphi(z_3)}{2}} \Big\rangle
\Big\langle
e^{i\left(q_1 - \sqrt{\frac{k}{2}} w_1 \right) W} \nonumber \\
\left(S^\pm V_{j_1}^{w_1} \right)_{j_1-\frac{1}{2}, m_1}(z_1)
e^{i\left(q_2 + \sqrt{\frac{k}{2}} w_2 \right) W}
V_{j_2,m_2}^{w_2}(z_2) e^{i\left(q_3 - \sqrt{\frac{k}{2}} w_3
\right) W} \left(S^\pm V_{j_3}^{w_3} \right)_{j_3-\frac{1}{2},
m_3}(z_3) \Big\rangle \nonumber
\end{eqnarray}

The conservation law for the field $H_2$ gives nonvanishing results
from terms containing either $S^{\pm}_{-\frac{1}{2}}$ and
 $S^{\pm}_{\frac{1}{2}}$ or
$S^{\pm}_{\pm\frac{1}{2}}$ and
$S^{\mp}_{\pm\frac{1}{2}}$.
Therefore if both vertices
 $\mathcal{Y}$ are of the same type ($+$ or $-$) one gets
\begin{eqnarray}
\Big\langle \mathcal{Y}_{j_1,m_1,q_1}^{-\frac{1}{2},w_1, (\pm)} \,
\mathcal{V}_{j_2,m_2,q_2}^{-1,w_2} \,
\mathcal{Y}_{j_3,m_3,q_3}^{-\frac{1}{2},w_3, (\pm)}
\Big\rangle = \Big\langle e^{i\left(q_1 - \sqrt{\frac{k}{2}}
w_1 \pm \frac{1}{\alpha'_+} \right) W} e^{i\left(q_2 +
\sqrt{\frac{k}{2}} w_2 \right) W} e^{i\left(q_3 -
\sqrt{\frac{k}{2}} w_3 \pm \frac{1}{\alpha'_+} \right) W}
\Big\rangle \quad \nonumber\\
\left\{ \Big\langle e^{\mp\frac{i}{2} H_2} e^{\pm\frac{i}{2} H_2}
\Big\rangle \Big\langle e^{i \left( + \frac{1}{2} + w_1 \right)
H_1} e^{i w_2 H_1} e^{i \left( - \frac{1}{2} + w_3 \right) H_1}
\Big\rangle \Big\langle V'_{j_1,m_1-\frac{1}{2},w_1}
V'_{j_2,m_2,w_2} V'_{j_3,m_3+\frac{1}{2},w_3} \Big\rangle
\right.\nonumber\\
\left. +\Big\langle e^{\pm\frac{i}{2} H_2} e^{\mp\frac{i}{2} H_2}
\Big\rangle \Big\langle e^{i \left( - \frac{1}{2} + w_1 \right)
H_1} e^{i w_2 H_1} e^{i \left( + \frac{1}{2} + w_3 \right) H_1}
\Big\rangle \Big\langle V'_{j_1,m_1+\frac{1}{2},w_1}
V'_{j_2,m_2,w_2} V'_{j_3,m_3-\frac{1}{2},w_3} \Big\rangle
\right\}\nonumber
\end{eqnarray}

\noindent
with the following conservation laws
\begin{eqnarray}
& &q_1 + q_2 + q_3 + \sqrt{\frac{k}{2}} \left(-w_1 + w_2 -
w_3\right)
\pm \sqrt{\frac{2}{k}} = 0 \label{leyqcRYVY}\\
& &w_1 + w_2 + w_3 = 0 \quad . \label{leywcRYVY}
\end{eqnarray}

It is interesting that these correlators cannot violate winding number
conservation. Moreover this condition arises from the conservation law of
the field $H_2$ which bosonizes the fermions
 $\psi^3$  and $\chi$. Thus the $U(1)$ factor determines which correlators
are nonvanishing.

Finally, if conditions
(\ref{leyqcRYVY}) and
(\ref{leywcRYVY}) are verified, the correlator becomes
\begin{eqnarray}
\Big\langle \mathcal{Y}_{j_1,m_1,q_1}^{-\frac{1}{2},w_1, (\pm)} \,
\mathcal{V}_{j_2,m_2,q_2}^{-1,w_2} \,
\mathcal{Y}_{j_3,m_3,q_3}^{-\frac{1}{2},w_3, (\pm)}
\Big\rangle = \left\{ \Big\langle
V'_{j_1,m_1-\frac{1}{2},w_1} V'_{j_2,m_2,w_2}
V'_{j_3,m_3+\frac{1}{2},w_3} \Big\rangle \right. \nonumber\\
\left. + \Big\langle V'_{j_1,m_1+\frac{1}{2},w_1} V'_{j_2,m_2,w_2}
V'_{j_3,m_3-\frac{1}{2},w_3} \Big\rangle \right\} \quad
\label{cRYVY3m}
\end{eqnarray}

If the internal vertex $\mathcal{V}$ is annihilated by $J_0^+$,
one can insert such operator and rewrite the correlators as
\begin{eqnarray}
\Big\langle \mathcal{Y}_{j_1,m_1,q_1}^{-\frac{1}{2},w_1, (\pm)} \,
\mathcal{V}_{j_2,m_2,q_2}^{-1,w_2} \,
\mathcal{Y}_{j_3,m_3,q_3}^{-\frac{1}{2},w_3, (\pm)}
\Big\rangle_{S^2} = \frac{j_3-j_1+m_1-m_3}{m_1-\frac{1}{2}-j_1}
\nonumber\\
\Big\langle V'_{j_1,m_1-\frac{1}{2},w_1} V'_{j_2,m_2,w_2}
V'_{j_3,m_3+\frac{1}{2},w_3} \Big\rangle \, \label{cRYVY4m}
\end{eqnarray}

There is an equivalent expression if  the internal operator
corresponds to a lowest weight state.

The case where the operators $\mathcal{Y}$ are not of the same type
($i.e.$ one is $+$ and the other one is  $-$)
can be treated similarly. The novelty is that in this case the correlation
function must violate $w$ conservation. The conservation laws imply that
each one of the terms in the correlator must satisfy different
conditions, namely
\begin{eqnarray}
\left(S_{\frac{1}{2}}^+ S_{\frac{1}{2}}^-\right) &:&
w_1 + w_2
+ w_3 + 1 = 0 \label
{leywcRYVYd1}
 \\
\left(S_{-\frac{1}{2}}^+ S_{-\frac{1}{2}}^-\right) &:&
w_1 + w_2 + w_3 - 1 = 0 \label
{leywcRYVYd2}
\end{eqnarray}

\noindent and besides, in both cases,
\begin{eqnarray}
q_1 + q_2 + q_3 + \sqrt{\frac{k}{2}} \left(-w_1 + w_2 - w_3\right)
= 0 \label{leyqcRYVYd}
\end{eqnarray}

Proceeding as before one obtains for the first condition (\ref{leywcRYVYd1})
\begin{eqnarray}
\Big\langle \mathcal{Y}_{j_1,m_1,q_1}^{-\frac{1}{2},w_1, (\pm)} \,
\mathcal{V}_{j_2,m_2,q_2}^{-1,w_2} \,
\mathcal{Y}_{j_3,m_3,q_3}^{-\frac{1}{2},w_3, (\mp)}
\Big\rangle = \Big\langle V'_{j_1,m_1-\frac{1}{2},w_1}
\widetilde{V'}^+_{j_2,m_2,w_2} V'_{j_3,m_3-\frac{1}{2},w_3}
\Big\rangle \quad \label{cRYVYd11}
\end{eqnarray}

\noindent whereas for the second one (\ref{leywcRYVYd2}) we find
\begin{eqnarray}
\Big\langle \mathcal{Y}_{j_1,m_1,q_1}^{-\frac{1}{2},w_1, (\pm)} \,
\mathcal{V}_{j_2,m_2,q_2}^{-1,w_2} \,
\mathcal{Y}_{j_3,m_3,q_3}^{-\frac{1}{2},w_3, (\mp)}
\Big\rangle = \Big\langle V'_{j_1,m_1+\frac{1}{2},w_1}
\widetilde{V'}^-_{j_2,m_2,w_2} V'_{j_3,m_3+\frac{1}{2},w_3}
\Big\rangle \quad . \label{cRYVYd11b}
\end{eqnarray}

The last correlator we shall consider is
$\langle \mathcal{Y} \mathcal{W} \mathcal{Y} \rangle$.
This case is interesting since there is a direct contribution from the
field $H_2$. The pattern of winding (non)
conservation is  contrary to the previous one, $i.e.$ $w$ conservation is
violated in the case where the vertices are both of type $+$ or of type
$-$ whereas it is conserved if they are of opposite type.

Unlike in the NS sector here the term containing
$\psi^3$ also contributes. It is convenient to rewrite the bosonization
(\ref{bosonpsichi}) as

\begin{equation} \label{bosH2cp}
\chi \pm \psi^3 = \sqrt{2} e^{\pm i H_2} \quad .
\end{equation}

Thus we consider
\begin{eqnarray}
\Big\langle \mathcal{Y}_{j_1,m_1,q_1}^{-\frac{1}{2},w_1, (\pm)} \,
\mathcal{O}_{j_2,m_2,q_2}^{-1,w_2} \,
\mathcal{Y}_{j_3,m_3,q_3}^{-\frac{1}{2},w_3, (\pm)}
\Big\rangle \label{cRYWY1}
\end{eqnarray}
\noindent where
\begin{equation} \label{cROW2}
\mathcal{O}_{j_2,m_2,q_2}^{-1,w_2} = \sqrt{2} e^{-\varphi} e^{i
\left( q_2 - \sqrt{\frac{k}{2}} w_2 \right)} \left( e^{i H_2} -
e^{-i H_2} \right) e^{i w_2 H_1} V'_{j_2,m_2,w_2} \quad .
\end{equation}

The term $e^{iH_2}$ ($e^{-iH_2}$) selects factors containing
$S^\pm_{\pm\frac{1}{2}}$
($S^\pm_{\mp\frac{1}{2}}$). Some of them conserve winding and others have
total winding $\pm 1$, therefore all the possibilities are present in this
case. This is due to the field $H_2$. The explicit computation is similar
to the cases discussed previously. The reader can easily fill up the
details.

\section{Conclusions}

The original motivation of this work was
to extend the Coulomb gas formalism for string theory on $AdS_3$ to the
supersymmetric case. We would
like to stress that the formalism presented in section 3.2 was developed
constructively. This indicates that not only is it possible to extend the
bosonic formulation of \cite{gn2, gn3} but also that the supersymmetric
Coulomb gas formalism designed from scratch gives the proper extension.
This is very important to assure the uniqueness of the basic objects in
the business:  screening operators
and identities. Actually we have considered and discarded the possibility
of constructing new operators of this sort in the supersymmetric theory.

In particular, we argued against the existence of screening
operators in $w\ne 0$ sectors and consequently the formalism
naturally obeys the winding non-conservation pattern of the
bosonic theory shown by Maldacena and Ooguri in \cite{mo3}.
Moreover this general pattern is preserved in the supersymmetric
theory. However, it was found that, due to selection rules related
to the fermionic part of the theory, some channels are inhibited
in this case. Thus, the possibility of violating winding
number conservation is dependent on the excitation number of the
operators involved in the supersymmetric correlation functions.

The method was employed
to compute two and three point functions of physical states in
both Neveu-Schwarz and Ramond sectors.
Correlators of Neveu-Schwarz states in different winding sectors, both
obeying and
violating winding number conservation were presented.
We found, as expected, that the supersymmetric correlators can
be expressed in terms of the corresponding bosonic ones.

Furthermore we explicitly computed two and three point correlators in the
Ramond sector.
We  analyzed the structure of the pattern of
violation to
 winding conservation and stressed the important role played in this
matter by the conservation laws of the field $H_2$,
related to the
$U(1)$ factor of the theory.

Important problems remain. Above all the computation of four point
functions. Even though the method we have presented is only an
approximation, valid near the
boundary of spacetime, we expect that if this is a consistent model this
approach
will exhibit the factorization properties of a unitary
theory.

\acknowledgments

C.N. would like to thank Daniel L\'opez for collaboration in the
initial stages of this programme. This work is supported by grants
from CONICET (PIP98 0873) and Universidad de Buenos Aires (UBACyT
X805), Argentina.

\end{document}